\documentclass{article}
\usepackage[utf8]{inputenc}
\usepackage{amsmath}
\usepackage{amsfonts}
\usepackage{amssymb}
\usepackage{caption}
\usepackage{float}
\usepackage{graphicx}
\usepackage{tikz}
\usepackage{mathtools}
\usepackage[margin=0.9in]{geometry}
\usepackage{calrsfs}
\usepackage{bbm}
\usepackage{enumitem}

\DeclareMathAlphabet{\pazocal}{OMS}{zplm}{F}{n}
\DeclareMathAlphabet{\pazocal}{OMS}{zplm}{N}{n}
\DeclareMathAlphabet{\pazocal}{OMS}{zplm}{B}{n}
\DeclareMathAlphabet{\pazocal}{OMS}{zplm}{D}{n}
\newcommand\subpsim{\mathrel{%
  \ooalign{\raise0.2ex\hbox{$p$}\cr\hidewidth\raise-0.8ex\hbox{\scalebox{0.9}{$\sim$}}\hidewidth\cr}}}
\newcommand\subnsim{\mathrel{%
  \ooalign{\raise0.2ex\hbox{$n$}\cr\hidewidth\raise-0.8ex\hbox{\scalebox{0.9}{$\sim$}}\hidewidth\cr}}}
\usepackage{indentfirst}
\usepackage{dirtytalk}
\usepackage[utf8]{inputenc}
\newcommand{\mat}[1]{\mbox{\boldmath{$#1$}}} 
\usepackage[caption=false]{subfig}
\usepackage{listings}
\usepackage{fancyhdr}
\newtheorem{theorem}{Theorem}[section]
\usepackage{enumitem}
\usepackage{enumitem}
\usepackage{diagbox}
\usepackage{booktabs}
\usepackage{subfig}
\providecommand{\keywords}[1]
{
  \small	
  \textbf{Keywords} #1
}
\usepackage{authblk}

\title{Wavelet estimation of nonstationary spatial covariance function}

\author[1]{Yangyang Chen}
\author[1,*]{Pedro A. Morettin}

\author[2]{Ronaldo Dias}
\author[1]{Chang Chiann.}

\affil[1]{Department of Statistics, Institute of Mathematics and Statistics, University of São Paulo}
\affil[2]{Department of Statistics, Institute of Mathematics, Statisctics and Scientific Computing, University of Campinas}

\date{}

\begin{document}

\maketitle

\let\thefootnote\relax\footnote{*Corresponding author.}
\let\thefootnote\relax\footnote{E-mail address: pam@ime.usp.br (P. A. Morettin).}

\begin{abstract}
This work proposes a new procedure for estimating the non-stationary spatial covariance function for Spatial-Temporal Deformation. The proposed procedure is based on a monotonic function approach. The deformation functions are expanded as a linear combination of the wavelet basis. The estimate of the deformation guarantees an injective transformation. Such that two distinct locations in the geographic plane are not mapped into the same point in the deformation plane. Simulation studies have shown the effectiveness of this procedure. An application to historical daily maximum temperature records exemplifies the flexibility of the proposed methodology when dealing with real datasets.
\end{abstract}


\section{Introduction}

In geostatistics it is common to assume that the stochastic process is stationary, which means that the distribution does not change when shifted in the origin of the index set, and is isotropic, that is, the process is invariant under rotations around the origin. But in practice, the assumption of stationarity and isotropic are often difficult to hold in real applications; see for instance Guttorp et al. (1994), Schmidt and O'Hagan (2003), and Finley (2011).

If a bijective transformation $f$ that maps the sampling locations at a geographic domain (G plane) into space representations of the deformation domain (D plane) is built, the spatial correlation can be considered isotropic in the D plan.
The injectivity of transformation is one of the most important requirements to guarantee that two distinct locations in the G plane are not mapped into the same point in the D plane. 
(See Damian et al. (2001)).
A sufficient condition for $f$ being injective is that the Jacobian determinant of $f$ is non-zero. 


To guarantee the injectivity of the mapping function, Choi and Lee (2000) suggested the box constraints for uniform cubic $B$-$spline$ deformation coefficients. 
Musse et al. (2001) enforced Jacobian positivity with a novel constrained hierarchical parametric model.
Chun and Fessler (2009) provided sufficient conditions that restrict $B$-$splines$ based deformation coefficients to ensure that the Jacobian of such transformation is positive which extended the conditions of Kim (2004). 
Although these methods ensure the positivity of Jacobian, deterministic models were used. Sampson and Guttorp (1992) introduced the topic with a stochastic model but no guarantee that the transformation based on thin-plate splines is injective. Damian et al. (2001) suggested a solution to guarantee the injectivity of the transformation in a stochastic model using a Bayesian approach. 

In this paper, we propose a method for nonstationary covariance function modeling, the deformation guarantees the injectivity of the transformation and is based on the monotonic function approach. Note that the wavelet expansion is also used for the deformation.

The plan of the paper is as follows. In section 2 we briefly review wavelets and introduce the spatio-temporal model used in the paper. Then, the deformation proposed is presented in section 3. Some simulations are described in section 4 and an application to historical daily maximum temperature records is given in section 5. Finally, the paper ends with some comments in section 6.


\section{Background}

\subsection{Wavelets}

In wavelets analysis, a function $f \in L^2(\mathbb{R})$ can be approximated by a linear combination of binary dilations $2^j$ and dyadic translations $k2^{-j}$ of a function $\phi(t)$, called scaling function or father wavelet (which is used for capturing the smooth and the low-frequency of the data) and/or of a function $\psi(t)$, called mother wavelet (which is used for capturing the details and the high-frequency of the data). Thus, a wavelet basis is composed of functions $\{ \phi_{j,k}(t) \cup \psi_{j,k}(t), j,k \in \mathbb{Z} \} \in L^2(\mathbb{R})$, where
\begin{equation}
    \phi_{j,k}(t) = 2^{j/2} \phi(2^j t-k),
\end{equation}
and
\begin{equation}
    \psi_{j,k}(t) = 2^{j/2} \psi(2^j t-k).
\end{equation}

One way of obtaining the scaling function is a solution of the equation
\begin{equation}
    \phi(t) = \sqrt{2} \sum_k l_k \phi(2t-k),
\end{equation}
and $\psi(t)$ is obtained from $\phi(t)$ by
\begin{equation}
    \psi(t) = \sqrt{2} \sum_k h_k \phi(2t-k),
\end{equation}
where
\begin{equation}
    h_k = (-1)^k l_{1-k}
\end{equation}
and
\begin{equation}
    l_k = \sqrt{2} \int_{-\infty}^{\infty} \phi(t)\phi(2t-k) dt.
\end{equation}

In multiresolution expansion, any function $f \in L^2(\mathbb{R})$ can be represented as 
\begin{equation}
f(t) = \sum_k c_{J_0,k}\phi_{J_0,k}(t) + \sum_{j\ge J_0}\sum_k d_{j,k} \psi_{j,k}(t),
\end{equation}
for some coarse scale $J_0$, usually taken as zero, for more details see Hardle et al. (1997).

There are many families of wavelets, such as the Mexican hat wavelet, Shannon wavelet, Daubechies wavelet, and Haar wavelet, the oldest and the simplest possible wavelet. 
In this paper, the well-known Mexican hat and Shannon wavelets are employed, they are continuous wavelets and are defined by analytical expressions. Figures 1 and 2 present the scaling ($\phi(t)$) and wavelet ($\psi(t)$) functions from these wavelets.

\begin{figure}[h]
    \centering
    \includegraphics[width=7cm]{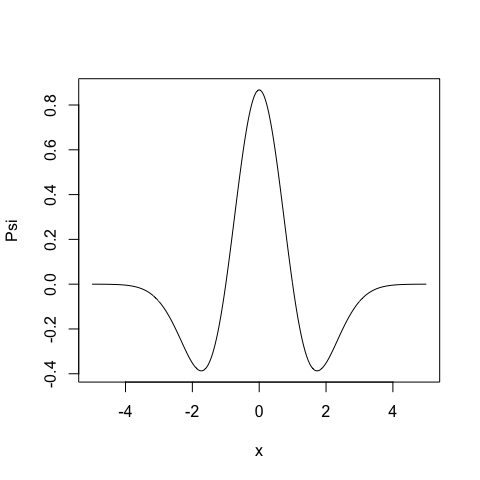}
    \caption{Wavelet function from the Mexican hat wavelet.}
\end{figure}

\begin{figure}[h!]
\centering
    \begin{tabular}{ll}
        \includegraphics[width=7cm]{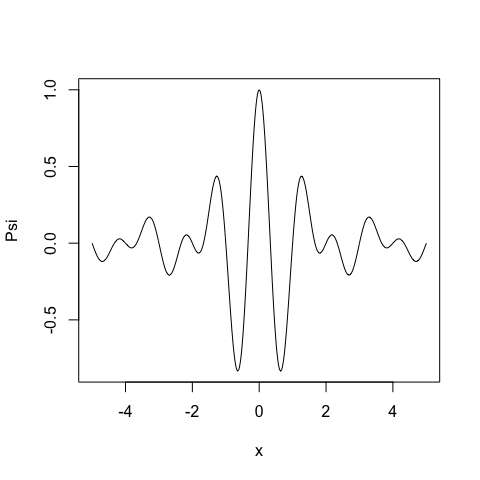}
        &
        \includegraphics[width=7cm]{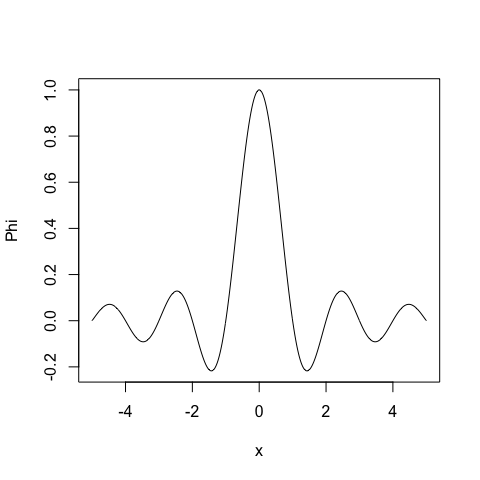}
    \end{tabular}
    \caption{Scaling (left) and wavelet (right) functions from the Shannon wavelet.}
\end{figure}

\subsection{Spatio-Temporal Model}

As the name suggests, spatio-temporal data are collected across space and time. Many fields, including geoscience, meteorology, neuroscience, and climate science, generate data that have both spatial and temporal components. For a comprehensive overview of spatio-temporal data, see Cressie and Wikle (2011). Let $Z_i(t) = Z(\textbf{x}_i, t)$, $i = 1,...,n$, $t = 1,...,T$, be a sample from a spatio-temporal process and the model proposed by Damian et al. (2001) will be considered for the underlying process:
\begin{equation}
    Z(\textbf{x}_i,t) = \mu(\textbf{x}_i,t) + \sqrt{\nu(\textbf{x}_i)}E_{\tau}(\textbf{x}_i) + E_{\epsilon}(\textbf{x}_i,t),
\end{equation}
where $\textbf{x}_i$ denotes location and $t$ time, $\mu(\textbf{x}_i, t)$ represents the spatio-temporal mean, $\nu(\textbf{x}_i)$ refers to the variance of the process observed at location $\textbf{x}_i$. $E_{\tau}(\textbf{x}_i)$ is a Gaussian spatial process and Cov[$E_{\tau}(\textbf{x}_i)$,$E_{\tau}(\textbf{x}_j)$] $\rightarrow$ 1 when $\textbf{x}_i \rightarrow \textbf{x}_j$ and $E_{\epsilon}$ represents measurement error and small-scale spatial variability. 

We assume that $\mu(\textbf{x}_i, t) \equiv \mu(\textbf{x}_i)$ is constant in time. Then, the covariance of the process at the locations $\textbf{x}_i$ and $\textbf{x}_j$, and time $t$, is given by
\begin{equation}
    \hbox{Cov}[Z(\textbf{x}_i,t), Z(\textbf{x}_j,t)] = \left\{\begin{array}{rll}
    \sqrt{\nu(\textbf{x}_i)\nu(\textbf{x}_j)}\hbox{Corr}[E_{\tau}(\textbf{x}_i)E_{\tau}(\textbf{x}_j)], & \hbox{if} & \textbf{x}_i \neq \textbf{x}_j, \\
    \nu(\textbf{x}_i)+\sigma_{\epsilon}^2, & \hbox{if} & \textbf{x}_i = \textbf{x}_j. \end{array}\right.
\end{equation}

Following Sampson and Guttorp (1992), the non-stationary spatial correlation of $E_{\tau}$ is obtained by:
\begin{equation}
    \hbox{Corr}[E_{\tau}(\textbf{x}_i)E_{\tau}(\textbf{x}_j)] = \rho(| f(\textbf{x}_i) - f(\textbf{x}_j) |),
\end{equation}
where $\rho$ is a correlation function with unknown parameter(s) and $f$ is a transformation that maps the sampling locations at the G plane into space representations of the D plane, where the spatial correlation can be considered isotropic.

\section{Deformation based on Monotonic Functions}

\subsection{Strictly Monotonic functions}

A monotonic function is a function that is either entirely non-increasing or non-decreasing. An important property of this kind of function is that if a function is a strictly monotonic function, then it is injective on its domain. In this paper, we use the monotonic function introduced by Ramsay (1998). It is an arbitrary twice differentiable strictly monotonic function defined on an interval closed on the left.

The strictly monotonic function $g$ satisfies the conditions: ln($Dg$) is differentiable and $D\{\ln (Dg)\} = D^2g/Dg$ is square Lebesgue integrable, and $D$ refers to the operation of taking the derivative and $D^{-1}$ means the integration operator. These conditions guarantee the function's first derivative is smooth and bounded almost everywhere. The following theorem shows that $g$ is a general solution of a differential equation


\begin{theorem}
    \normalfont (Ramsay (1998)) 
    Every monotonic function $g$ is representable as either 
    \begin{equation}
        g(x) = C_0 + C_1 D^{-1} \{ \exp(D^{-1} w(x)) \}
    \end{equation}
    or as a solution of the homogeneous linear differential equation
    \begin{equation}
        D^2g(x) = w(x) Dg(x),
    \end{equation}
    where $w(x)$ is a Lebesgue square integrable function and $C_0$ and $C_1$ are arbitrary constants.
\end{theorem}

Note that the function $g(x)$ is strictly monotone increasing, $Dg(x) = e^W(x)$, where $W(x) = D^{-1}w(x) +\log C_1$ were assumed, for more details see Ramsay and Silverman (2006).

\subsection{Basis function expansions}

By Theorem 3.1, instead of estimating the constrained function $g$, the problem becomes computing the unconstrained function $w$. Since $w$ can be positive or negative, we expand it as a linear combination of a set of wavelet basis functions.

As previously mentioned, two types of wavelets will be used:
\begin{itemize}
    \item Mexican hat, given by
    \begin{equation}
        \psi^{Mex}(x) = \frac{2}{\sqrt{3}\pi^{1/4}}(1-x^2) e^{-x^2/2}.
    \end{equation}
    Then we can expand $\omega$ as
    \begin{equation}
        \omega_{Mex}(x) = \sum_{j=0}^{J} \sum_{k=0}^{2^j-1}c_{j,k} \psi^{Mex}_{j,k}(x).
    \end{equation}
    \item Shannon wavelet, given by
    \begin{eqnarray}
        \psi^{Shan}(x) &=& \hbox{sinc}(\frac{x}{2}) \cos (\frac{3\pi x}{2}) = 2 \hbox{sinc}(2x) - \hbox{sinc}(x), \\
        \phi^{Shan}(x) &=& \hbox{sinc}(x),
    \end{eqnarray}
    where $\hbox{sinc}(x) = \frac{\sin \pi x}{\pi x}$. And $\omega$ can be expressed by
    \begin{equation}
        \omega_{Shan}(x) = c_0\phi^{Shan}(x) + \sum_{j=0}^{J} \sum_{k=0}^{2^j-1}c_{j,k} \psi^{Shan}_{j,k}(x).
    \end{equation}
\end{itemize}

\subsection{Deformation based on Monotonic Functions}

Let $\textbf{x}_i = (x_{i1},x_{i2})$, location $i$ in a G plane and $\textbf{y}_i = (y_{i1},y_{i2})$, its deformed location in a D plane, $i = 1,...,n$. As there is no natural sort order in $\mathbb{R}^2$, it's very difficult to get a bidimensional monotonic function. Adapting the generalized additive model introduced by Hastie e Tibshirani (1990) to estimate the deformation, we can consider each coordinate of the representation of the D plane as the response variable and the coordinates of the location in the G plane as predictor variables and thus we have an additive model as
\begin{equation}
    y_{il} = \beta_0 + \sum_{j=1}^2 g^l_j(x_{ij}) + \epsilon_{il}, \ l = 1, 2,
\end{equation}
where $\beta_0$ represent intercept and $\epsilon_{il}$ indicates random effect.

Suppose that $\beta_0 = 0$ and random effect is null.
Therefore, the representation in the D plane, $\textbf{y}_i = (y_{i1},y_{i2})$, can be written as
\begin{equation}
    \textbf{y}_i = \left[ \begin{array}{c} y_{i1}\\y_{i2}\\
    \end{array}\right]  = 
    \left[ \begin{array}{c} 
    g^1_1(x_{i1}) + g^1_2(x_{i2}) \\
    g^2_1(x_{i1}) + g^2_2(x_{i2})\\
    \end{array}\right],
\end{equation}
where $g_1^l(x_{i1})$ and $g_2^l(x_{i2})$ are monotonic functions given by (11). Since $g^l_1(x_{i1})$ and $g^l_2(x_{i2})$ are strictly monotonic on the range $[0, \infty)$, $y_{il}$ is also strictly monotonic in this range. Then, $y_{il}$ is a injective function. And we can write 
\begin{equation}
    y_{il} = C^l_{10} + C^l_{11}D^{-1}\exp\{ D^{-1} \omega^{l1}(x_{i1}) \} + C^l_{20} + C^l_{21}D^{-1} \exp\{D^{-1} \omega^{l2}(x_{i2}) \}.
\end{equation}


According to Theorem 3.1, $C_0$ and $C_1$ are arbitrary constants, suppose that $C^l_{10} = C^l_{20} = 0$ and $C^l_{11} = C^l_{21} = 1$. Therefore, (20) can be simplified to
\begin{equation}
    y_{il} = D^{-1} \exp\{D^{-1} \omega^{l1}(x_{i1}) \} +D^{-1}\exp\{ D^{-1} \omega^{l2}(x_{i2}) \}.
\end{equation}

Note that $\omega^{l1}(x_{i1})$ and $\omega^{l2}(x_{i2})$ can be written in the form (14) or (17). Thus, the estimated deformations using Mexican hat and Shannon wavelets can be written as
\begin{eqnarray}
    \hat{y}^{Mex}_{il} = D^{-1} \exp\{ \omega^{l1}_{Mex}(x_{i1}) \} +D^{-1} \exp\{ \omega^{l2}_{Mex}(x_{i2}) \}, \\
    \hat{y}^{Shan}_{il} = D^{-1}\exp\{ \omega^{l1}_{Shan}(x_{i1}) \} +D^{-1} \exp\{ \omega^{l2}_{Shan}(x_{i2}) \},
\end{eqnarray}
for $i = 1,...,n, \ l = 1, 2$.

\subsection{Process Optimization}

Let $\textbf{x}_i = (x_{i1}, x_{i2}) \in G \subset \mathbb{R}^2$ coordinates of location $i$ in a G plane and suppose that a collection of $n$ locations $((x_{11},x_{12}),...,(x_{n1},x_{n2}))$ was obtained. The optimization procedure is used to estimate the deformation $\textbf{y}_i$'s, the spatial variance $\nu = \nu(\textbf{x})$ and the parameters $\mat{\theta}$ of the correlation function $\rho$, maximizing the likelihood function of the samples $z_{it} = Z(\textbf{x}_i, t)$, $i = 1,..,n$, $t = 1,..,T$.

Assuming $\textbf{z}_t = (z_{1t},...,z_{nt})' \sim N_n(\mat{\mu}, \Sigma_{\mat{\eta}})$, the likelihood function is
\begin{eqnarray}
    L(\mat{\mu}, \Sigma_{\mat{\eta}}|\textbf{z}) &=& (2\pi)^{-nT/2} \times det(\Sigma_{\mat{\eta}})^{-T/2} \times \exp \{ -\frac{1}{2}\sum_{t=1}^{T}(\textbf{z}_t - \mat{\mu})^T \Sigma_{\mat{\eta}}^{-1} (\textbf{z}_t - \mat{\mu})\} \nonumber\\
    &=& (2\pi)^{-nT/2} \times det(\Sigma_{\mat{\eta}})^{-T/2} \times \exp \{ -\frac{1}{2} tr[\Sigma^{-1}_{\mat{\eta}} (\sum_{t=1}^T (\textbf{z}_t - \bar{\textbf{z}})(\textbf{z}_t - \bar{\textbf{z}})^T + T(\bar{\textbf{z}} - \mat{\mu})(\bar{\textbf{z}} - \mat{\mu})^T)] \} \nonumber\\
    &=& (2\pi)^{-nT/2} \times det(\Sigma_{\mat{\eta}})^{-T/2} \times \exp \{ -\frac{T}{2} tr[\Sigma_{\mat{\eta}}^{-1} S] - \frac{T}{2}(\bar{\textbf{z}} - \mat{\mu})^T \Sigma_{\mat{\eta}}^{-1} (\bar{\textbf{z}} - \mat{\mu})\},
\end{eqnarray}
where $\bar{\textbf{z}}$ is the vector of means at each location, $S$ is sample spatial covariance matrix whose element $S_{ij} = \sum_{t = 1}^{T} (z_{it} - \bar{z}_{i.})(z_{jt}-\bar{z}_{j.})$ and the covariance matrix
\begin{equation}
    \Sigma_{\mat{\eta} = ( \textbf{c}, \nu, \mat{\theta})} = (\sigma_{ij}),
\end{equation}
where 
\begin{equation}
    \sigma_{ij} = \sqrt{v_iv_j} \rho_{\mat{\theta}}( |\textbf{y}_i - \textbf{y}_j|),
\end{equation}
with the coordinates $\textbf{y}_i$ and $\textbf{y}_j$ of the representation of the D plane, the spatial variance $\nu_i$ and $\nu_j$ at locations $i$ and $j$, respectively, and the parameters $\mat{\theta}$ of the correlation function $\rho$.

Without loss of generality, set $\mat{\mu} = \textbf{0}$, the optimization problem becomes
\begin{equation}
    \underset{\mat{\eta}}{\hbox{max}} \ L(\Sigma_{\mat{\eta}}|\textbf{z}).
\end{equation}

The optimization procedure to estimate the covariance matrix (25) consists of the following steps:
\begin{itemize}
    \item[1.] Let $\textbf{c}^0$ be the initial values of \textbf{c}, coefficients of $\omega$ and calculate $\mat{\gamma}^0 = (\nu^0, \mat{\theta}^0)$ such that
    \begin{equation}
        \mat{\gamma}^0 = \hbox{arg max} \ l(\Sigma_{\nu, \mat{\theta}}|\textbf{z}, \textbf{c}^0), 
    \end{equation}
    where $l(\Sigma_{\nu, \mat{\theta}}|\textbf{z}, \textbf{c}^0) = \log L(\Sigma_{\nu, \mat{\theta}}|\textbf{z}, \textbf{c}^0)$.
    \item[2.] Given $\mat{\gamma}^0$ obtained in Step 1, calculate
    \begin{equation}
        \textbf{c}^1 = \hbox{arg max} \ l(\textbf{c} | \textbf{z}, \Sigma_{\nu^0, \mat{\theta}^0}),
    \end{equation}
    where $l(\textbf{c} | \textbf{z}, \Sigma_{\nu^0, \mat{\theta}^0}) = \log L(\textbf{c} | \textbf{z}, \Sigma_{\nu^0, \mat{\theta}^0})$.
    \item[3.] Replace $\textbf{c}^0$ by $\textbf{c}^1$, return to Step 1.
    \item[4.] Repeat Step (2) - (3) until convergence.
\end{itemize}

\section{Simulations}

This section presents some simulations in order to assess the performance of the proposed procedure. These simulations are carried out in the cases that the deformations are generated by functions linear, quadratic, non-linear and wavelet.

The deformed coordinates $\textbf{y}_i$'s, i.e. representations of D plane at location $i$, $i = 1,..,n$, are generated as follows:
\begin{itemize}
    \item Linear case:
        \begin{eqnarray}
        y_{i1} = 0.75x_{i1}+x_{i2}, \\
        y_{i2} = x_{i1}+0.25x_{i2}.
        \end{eqnarray}
    \item Quadratic case:
        \begin{eqnarray}
        y_{i1} = -0.5(x_{i1}-0.5)^2 + (x_{i2}-0.5) + 0.6, \\
        y_{i2} = (x_{i1}-0.5) - 0.5(x_{i2}-0.5)^2 + 0.6.
        \end{eqnarray}
    \item Non-linear case:
        \begin{eqnarray}
        y_{i1} = \cos(\hbox{angle})(x_{i1}-0.5) + \sin(\hbox{angle})(x_{i2}-0.5) + 0.5, \\
        y_{i2} = -\sin(\hbox{angle})(x_{i1}-0.5) + \cos(\hbox{angle})(x_{i2}-0.5) + 0.5,
        \end{eqnarray}
        where $\hbox{angle} = 2.5\exp \{ -(x_{i1}-0.5)^2-(x_{i2}-0.5)^2 \} + 3\pi/2$.
    \item Wavelet case:
        \begin{eqnarray}
        y_{i1} = D^{-1}\exp\{ \sum_{j=0}^1 \sum_{k=0}^{2^j-1} c^{11}_{j,k} \psi_{j,k}(x_{i1}) \}+ D^{-1}\exp\{ \sum_{j=0}^1 \sum_{k=0}^{2^j-1} c^{12}_{j,k} \psi_{j,k}(x_{i2})\}, \\
        y_{i2} = D^{-1}\exp\{ \sum_{j=0}^1 \sum_{k=0}^{2^j-1} c^{21}_{j,k} \psi_{j,k}(x_{i1})\}+ D^{-1}\exp\{ \sum_{j=0}^1 \sum_{k=0}^{2^j-1} c^{22}_{j,k} \psi_{j,k}(x_{i2})\},
        \end{eqnarray}
        where $\psi(t)$ are Mexican hat wavelets, $c^{11}_{0,0} = 0.25$, $c^{11}_{1,0} = 0.01$, $c^{11}_{1,1} = -0.036$, $c^{12}_{0,0} = -0.37$, $c^{12}_{1,0} = 0.065$, $c^{12}_{1,1} = -1.2$, $c^{21}_{0,0} = -0.032$, $c^{21}_{1,0} = -0.043$, $c^{21}_{1,1} = -1$, $c^{22}_{0,0} = -0.031$, $c^{22}_{1,0} = 0.11$, $c^{22}_{1,1} = 0.19$.
\end{itemize}


The simulations consist of the following steps:
\begin{itemize}
    \item[1.] Let $n$ = 50 sample locations generated in the geographical domain G = $[0, 1] \times [0, 1]$, the coordinates (longitude and latitude) are generated following a uniform dispersion in $(0, 1)$. Figure 3 shows the sampling locations, and the deformed locations are presented in Figure 4.
    \item[2.] The sample data (\textbf{z} = ($\textbf{z}_1, ..., \textbf{z}_T$)) are simulated from a Gaussian random field with mean function $\mu(\textbf{z}) = \textbf{0}$ and covariance function
    \begin{equation}
        \hbox{Cov}(\textbf{z}_i, \textbf{z}_j) = \nu \rho_{\mat{\theta}}(|\textbf{y}_i- \textbf{y}_j|) + \sigma_{\epsilon}^2 \mathbb{I}_{(\textbf{z}_i=\textbf{z}_j)} = \nu \exp \{-|\textbf{y}_i - \textbf{y}_j|/\theta \} + \sigma_{\epsilon}^2 \mathbb{I}_{(\textbf{z}_i=\textbf{z}_j)}
    \end{equation}
    with parameters $\nu = 1, \ \theta = 0.25$ and $\sigma_{\epsilon}^2 = 0.05$. The length of each series was fixed at $T = 2048$.
    \item[3.] Calculate the estimates of $\textbf{y}_i$, $\hat{\textbf{y}}_i = (\hat{y}_{i1}, \hat{y}_{i2})$ for $i = 1,..., n = 50$ and the parameters of the covariance function using the optimization procedure described in section 3.4 with Mexican hat and Shannon wavelets. 
    \item[4.] Calculate mean squared error (MSE) of the estimates of the correlation matrix,
    \begin{equation}
        \hbox{MSE} = \frac{1}{n^2} \sum_{i=1}^{n} \sum_{j=1}^{n} (\hbox{corr}_{ij} - \exp \{-|\hat{\textbf{y}}_i - \hat{\textbf{y}}_j|/\hat{\theta} \})^2,
    \end{equation}
    where $\hbox{corr}_{ij}$ is the element of the $i$th row and $j$th column in the correlation matrix of \textbf{z}, i.e. is the correlation between $\textbf{z}_i$ and $\textbf{z}_j$.
\end{itemize}

\begin{figure}[h]
    \centering
    \includegraphics[scale=0.35]{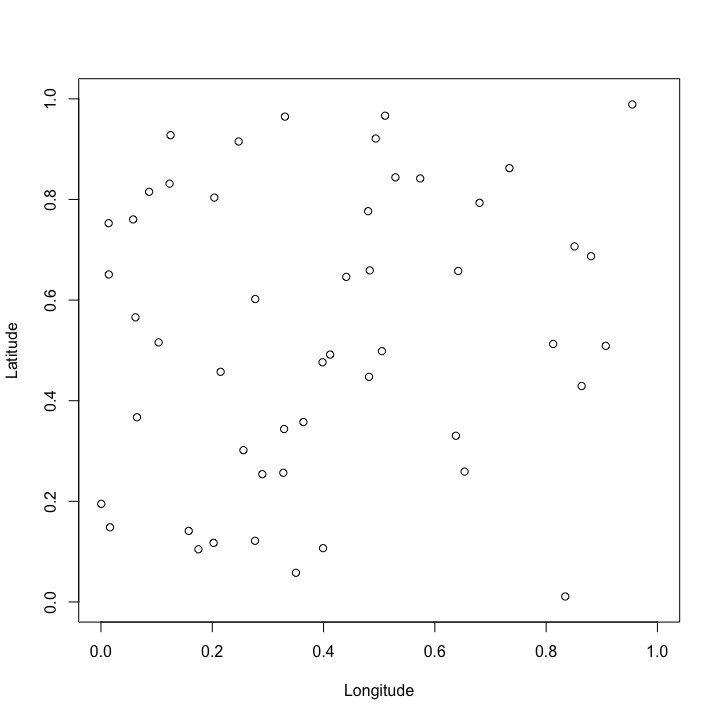}
    \caption{Simulated locations.}
    \label{fig:my_label}
\end{figure}

\begin{figure}[h!]
    \centering
    
    \subfloat[Linear Deformation]{\includegraphics[width=0.35\textwidth, keepaspectratio]{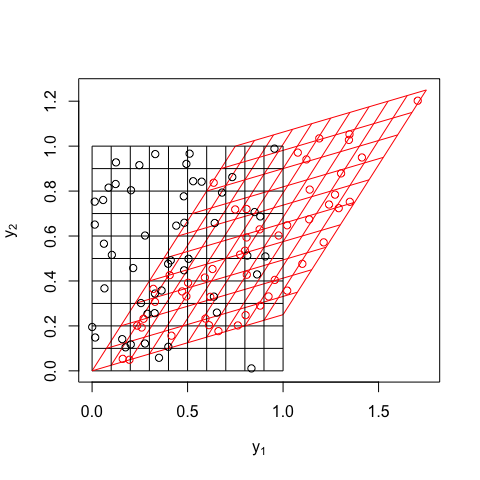}\label{fig:taba}}
    \subfloat[Quadratic Deformation]{\includegraphics[width=0.35\textwidth, keepaspectratio]{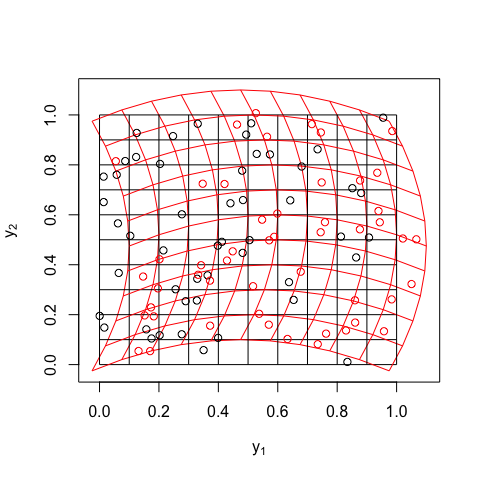}\label{fig:tabb}}\\
    \subfloat[Non-linear Deformation]{\includegraphics[width=0.35\textwidth, keepaspectratio]{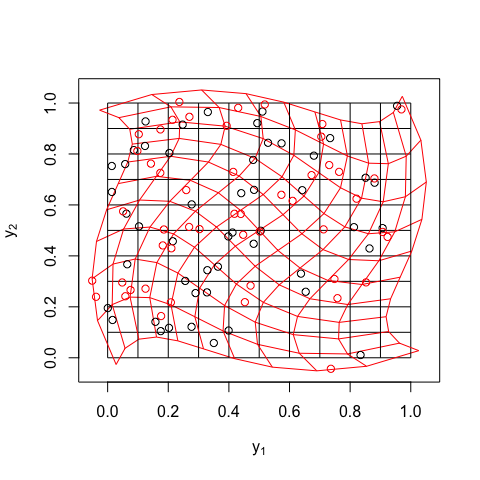}\label{fig:tabc}}
    \subfloat[Wavelet Deformation]{\includegraphics[width=0.35\textwidth, keepaspectratio]{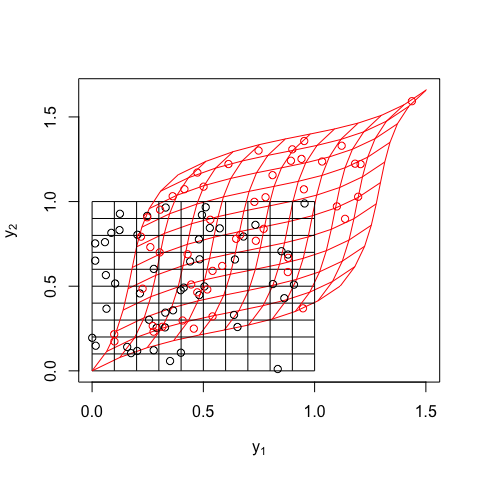}\label{fig:tabd}}
    \caption{Sampling locations (and regular grid) in G plane, in black, and deformed locations (and deformed grid) in plane D, in red.}
    \label{fig:my_label}
\end{figure}


Table 1 shows the estimated parameters of the covariance function and the MSE of the estimates of the correlation matrix in several fits. Observe that in the linear, non-linear, and wavelet case, the estimates that used $J$ = 3 and Shannon wavelet were closer to the true values of the parameters, whereas in the quadratic case, they were closer when using the Mexican hat wavelet.
Note that in three cases the lowest MSE was obtained using the Mexican hat wavelet, however with $J$ = 2 in the linear case and $J$ = 3 in the non-linear and wavelet case. Only in the quadratic case, the smallest MSE was obtained using the Shannon wavelet and $J$ = 4.

Figures 5-12 show the estimated deformation and the scatter plots of the upper-diagonal entries of the estimated correlation matrices for the sample data, versus the true correlation matrices. The comparison of estimated correlation matrices is of great convenience to evaluate the estimation results.
According to the MSE's in Table 1, the scatter plot is more accurate for smaller MSE. Note that in the linear and wavelet case, the estimated deformations are very close to the true one.

\clearpage

\begin{table}[H]
\centering
\caption{Estimated parameters of the covariance function and MSE of the correlation matrix of different fits.}
\footnotesize
\begin{tabular}{cccc}
    True value of the parameters: & $\nu$ = 1 & $\theta = 0.25$ & $\sigma^2_{\epsilon}$ = 0.05 \\
\end{tabular}
    \begin{tabular}{ccccccc}
    \toprule
     &\multicolumn{3}{c}{Mexican hat} & \multicolumn{3}{c}{Shannon} \\ 
     \bottomrule
     &\multicolumn{6}{c}{Linear Deformation} \\
     &J = 2 & J = 3 & J = 4 & J = 2 & J = 3 & J = 4 \\\hline
    $\nu$   & 1.03540 & 1.04229 & 1.03757 & 1.03493 & 1.02579 & 1.03192 \\   
    $\theta$   & 0.19197 & 0.18178  & 0.20000 & 0.20204 & 0.23975 & 0.21140\\ 
    $\sigma^2_{\epsilon}$  & 0.04117 & 0.03905  & 0.03771 & 0.04380 & 0.06739 & 0.04512\\
    MSE & 0.00173 & 0.00505 & 0.00290 & 0.00196 & 0.00409 & 0.00708\\
    \bottomrule
    \toprule
    &\multicolumn{6}{c}{Quadratic Deformation} \\
     &J = 3 & J = 4 & J = 5 & J = 3 & J = 4 & J = 5 \\\hline
    $\nu$  & 1.01631 & 1.03535 & 1.01537 & 1.04410 & 1.04491 & 1.00799 \\   
    $\theta$  & 0.25529 & 0.26601  & 0.25003 & 0.26182 & 0.25878 & 0.25973 \\ 
    $\sigma^2_{\epsilon}$  & 0.04272 & 0.04371  & 0.04082 & 0.04140 & 0.04032 & 0.04690 \\
    MSE & 0.00309 & 0.00165 & 0.00195 & 0.00220 & 0.00124 & 0.00205 \\
    \bottomrule
    \toprule
    &\multicolumn{6}{c}{Non-linear Deformation} \\
    & J = 3 & J = 4 & J = 5 & J = 3 & J = 4 & J = 5 \\\hline
    $\nu$  & 1.04400 & 1.04378 & 1.02342 & 1.02903 & 0.99906 & 1.04291 \\   
    $\theta$  & 0.22920 & 0.21801  & 0.23132 & 0.25214 & 0.24048 & 0.23826 \\ 
    $\sigma^2_{\epsilon}$  & 0.04117 & 0.03905  & 0.03771 & 0.04183 & 0.03259 & 0.03297 \\
    MSE & 0.00387 & 0.00539 & 0.00485 & 0.00409 & 0.00436 & 0.00451 \\
    \bottomrule
    \toprule
    &\multicolumn{6}{c}{Wavelet Deformation} \\
    & J = 3 & J = 4 & J = 5 & J = 3 & J = 4 & J = 5 \\\hline
    $\nu$  & 1.00827 & 1.02275 & 1.01732 & 1.02485 & 1.00019 & 1.00496 \\   
    $\theta$  & 0.19449 & 0.19605  & 0.20986 & 0.22260 & 0.20877 & 0.21827 \\ 
    $\sigma^2_{\epsilon}$  & 0.05629 & 0.04681  & 0.05375 & 0.06022 & 0.06107 & 0.06728 \\
    MSE & 0.00179 & 0.00279 & 0.00218 & 0.00296 & 0.00222 & 0.00226 \\
    \bottomrule
    \end{tabular}

\end{table}

\begin{figure}[h]
\centering
    \begin{tabular}{lll}
        \includegraphics[width=4.8cm]{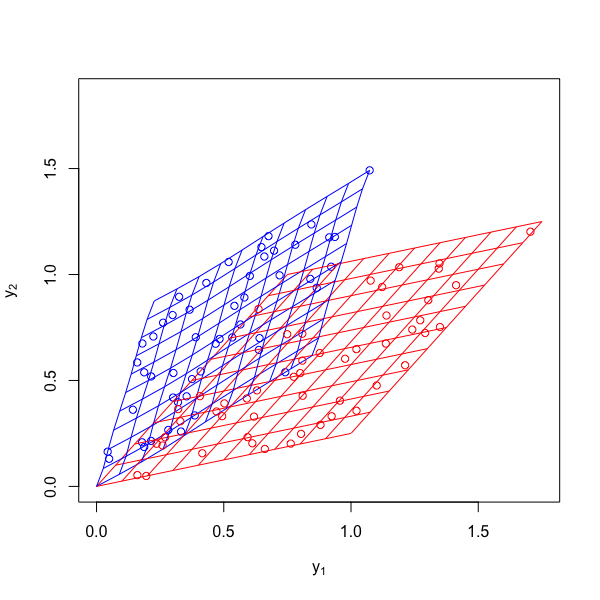}
        &
        \includegraphics[width=4.8cm]{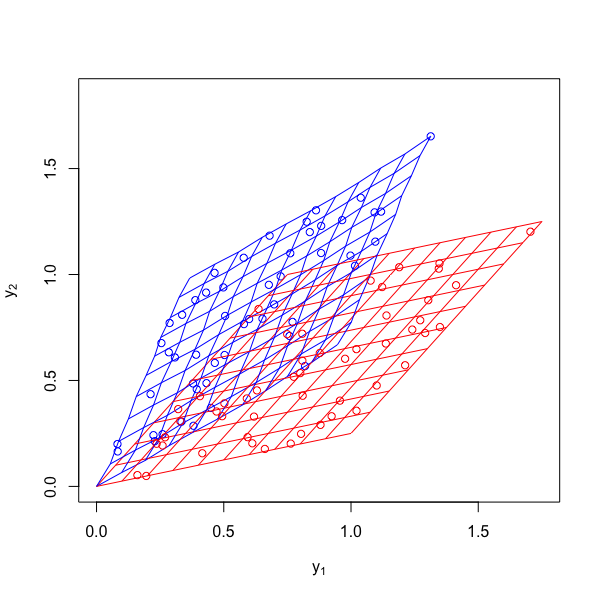}
        &
        \includegraphics[width=4.8cm]{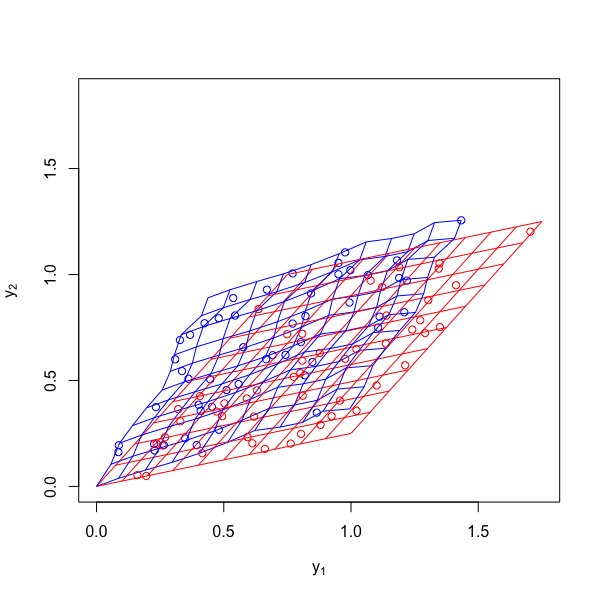}
        \\
        \includegraphics[width=4.8cm]{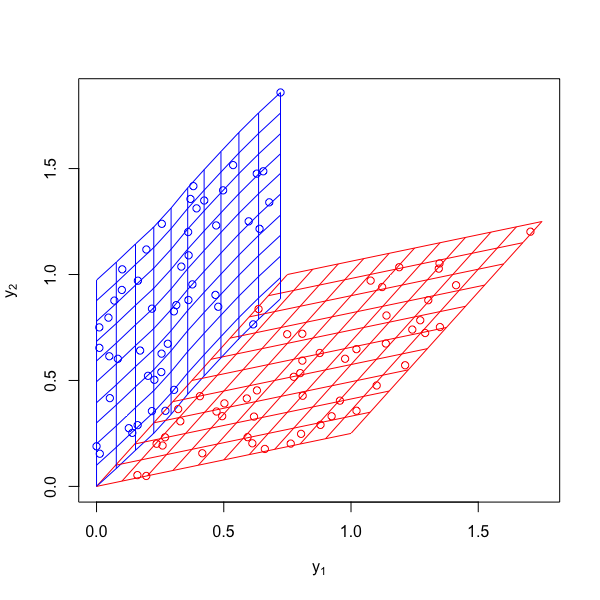}
        &
        \includegraphics[width=4.8cm]{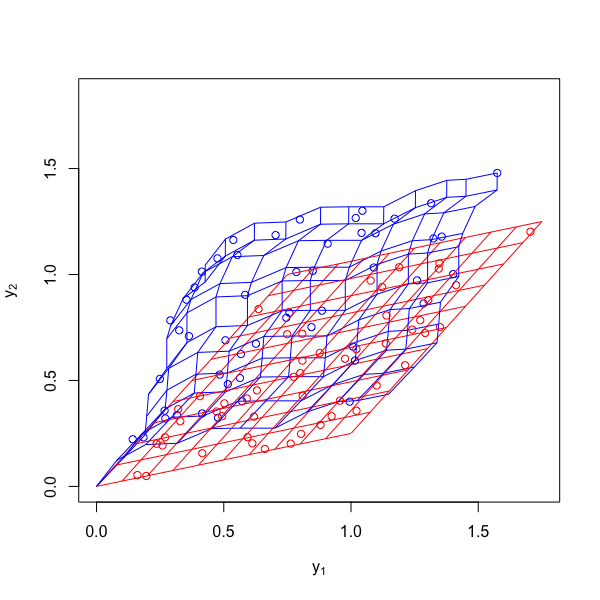}
        &
        \includegraphics[width=4.8cm]{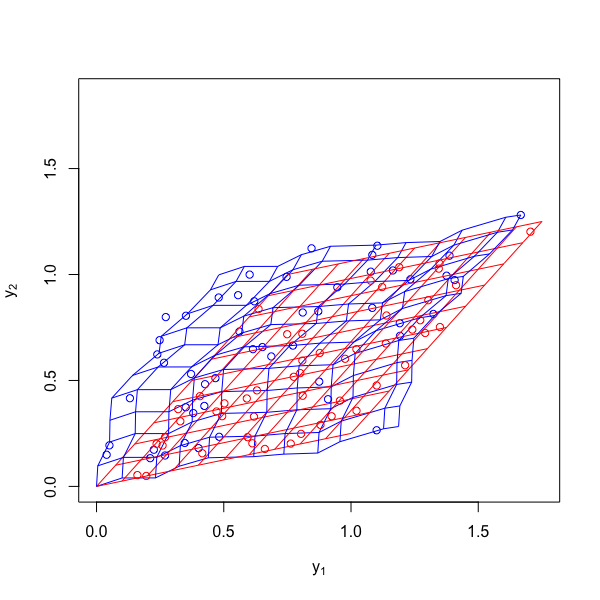}
    \end{tabular}
    \caption{Estimated deformation (in blue) when $J = 2, 3, 4$ (from left to right) for linear deformation case using Mexican hat (upper) and Shannon (bottom) wavelets.}
\end{figure}
\begin{figure}[H]
\centering
    \begin{tabular}{lll}
        \includegraphics[width=4.8cm]{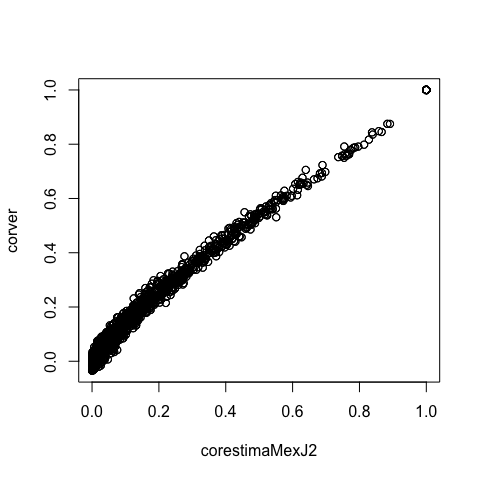}
        &
        \includegraphics[width=4.8cm]{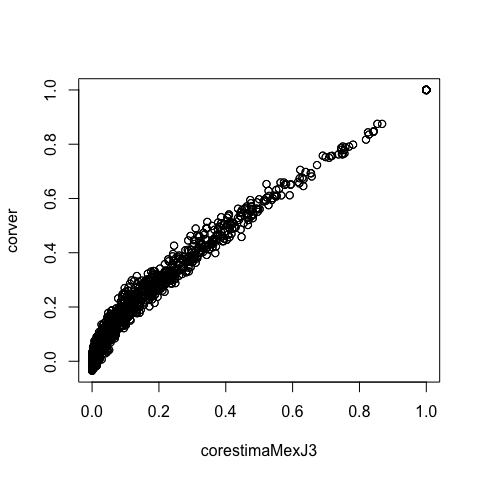}
        &
        \includegraphics[width=4.8cm]{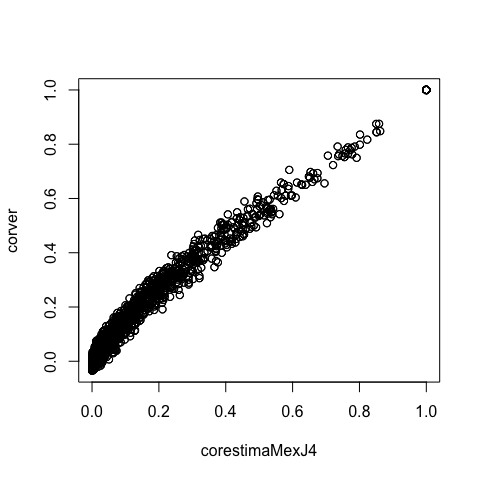}
        \\
        \includegraphics[width=4.8cm]{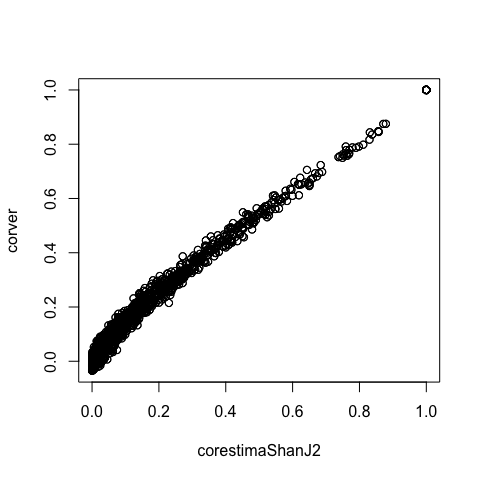}
        &
        \includegraphics[width=4.8cm]{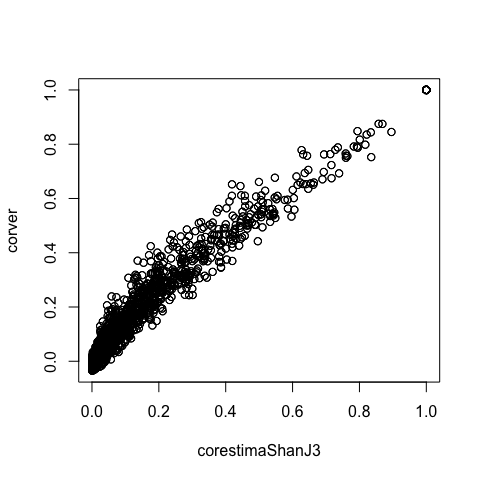}
        &
        \includegraphics[width=4.8cm]{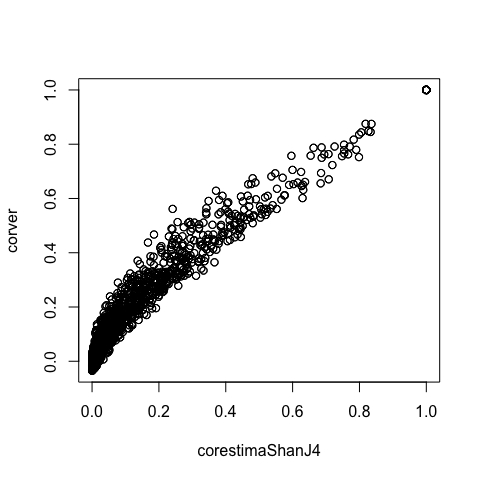}
    \end{tabular}
    \caption{Comparison of the estimated correlation matrix versus the true correlation matrix when $J = 2, 3, 4$ (from left to right) for linear deformation case using Mexican hat (upper) and Shannon (bottom) wavelets.}
\end{figure}

\begin{figure}[H]
\centering
    \begin{tabular}{lll}
        \includegraphics[width=4.8cm]{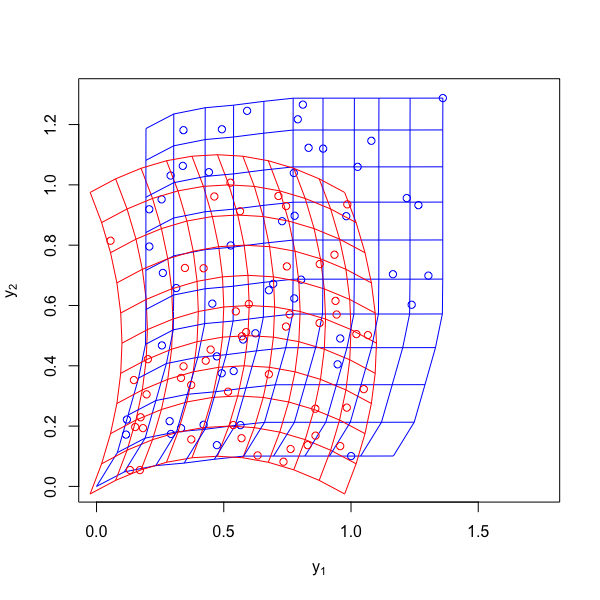}
        &
        \includegraphics[width=4.8cm]{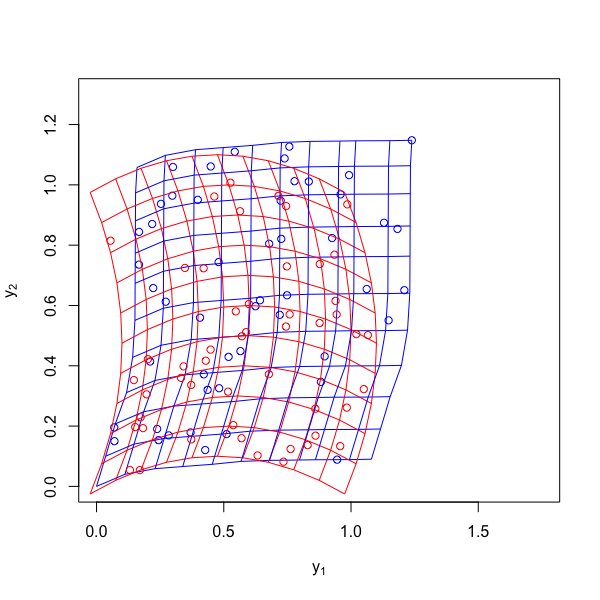}
        &
        \includegraphics[width=4.8cm]{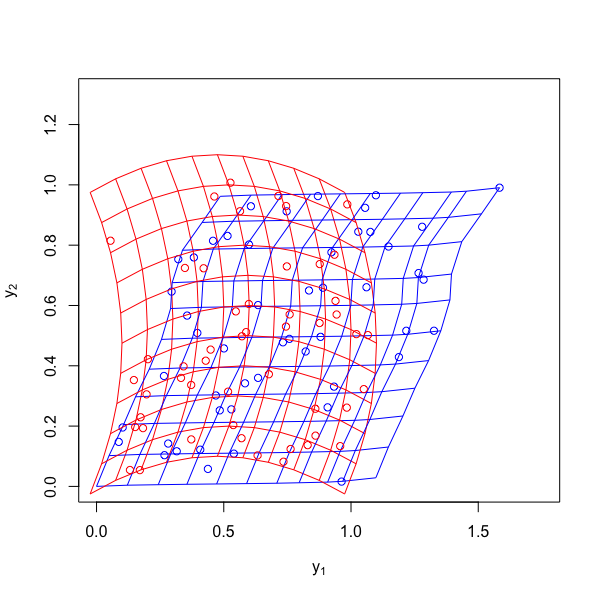}
        \\
        \includegraphics[width=4.8cm]{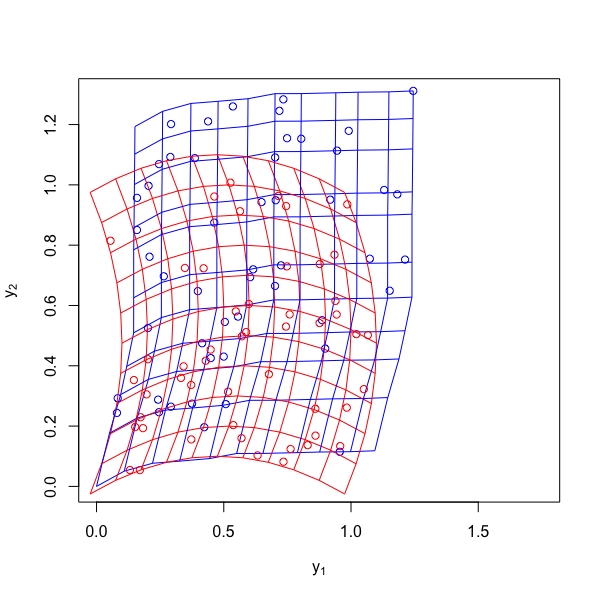}
        &
        \includegraphics[width=4.8cm]{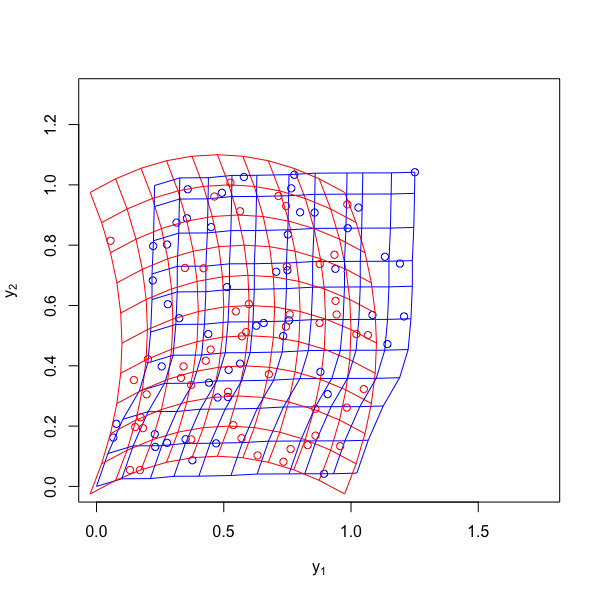}
        &
        \includegraphics[width=4.8cm]{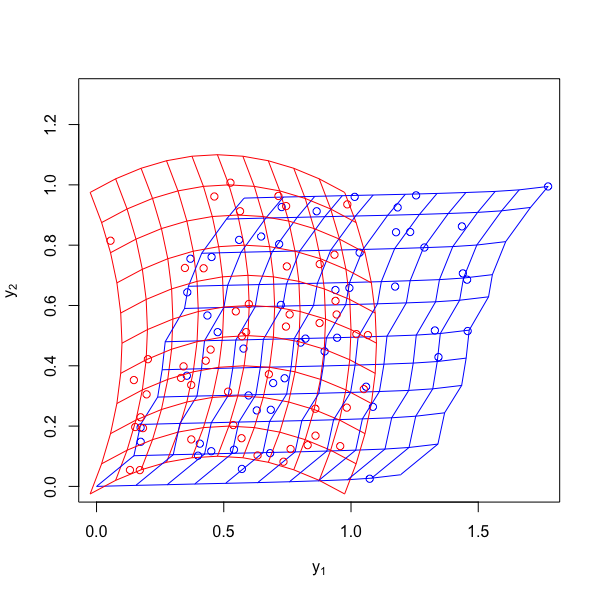}
    \end{tabular}
    \caption{Estimated deformation (in blue) when $J = 3, 4, 5$ (from left to right) for quadratic deformation case using Mexican hat (upper) and Shannon (bottom) wavelets.}
    \begin{tabular}{lll}
        \includegraphics[width=4.8cm]{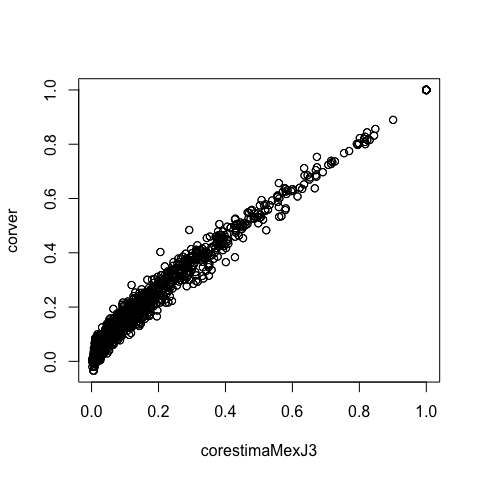}
        &
        \includegraphics[width=4.8cm]{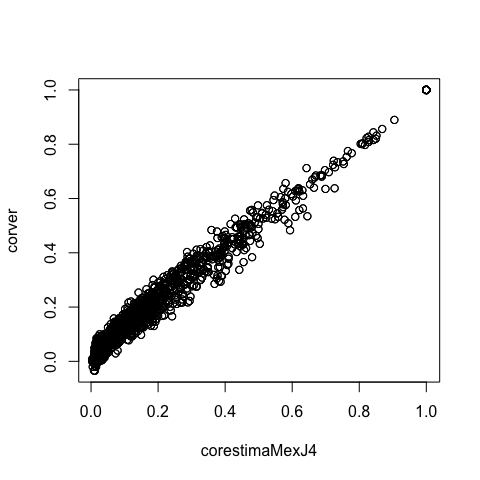}
        &
        \includegraphics[width=4.8cm]{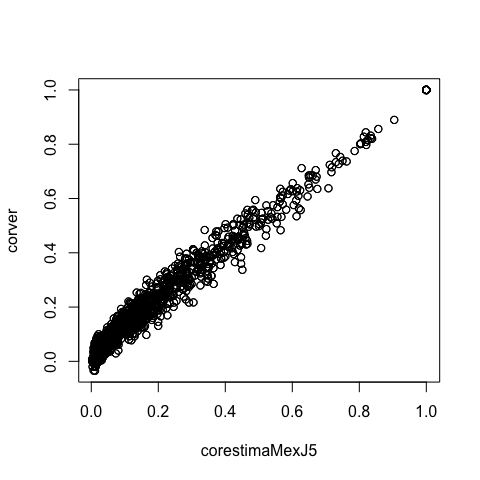}
        \\
        \includegraphics[width=4.8cm]{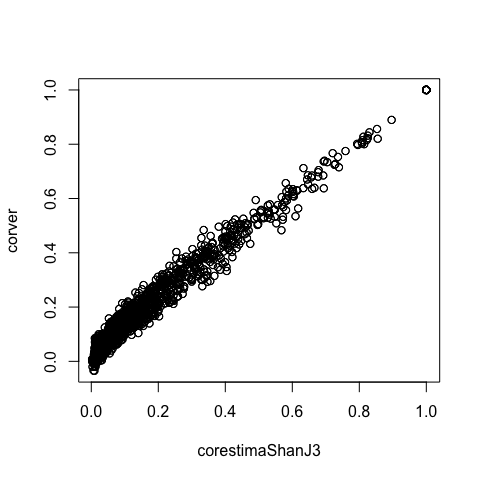}
        &
        \includegraphics[width=4.8cm]{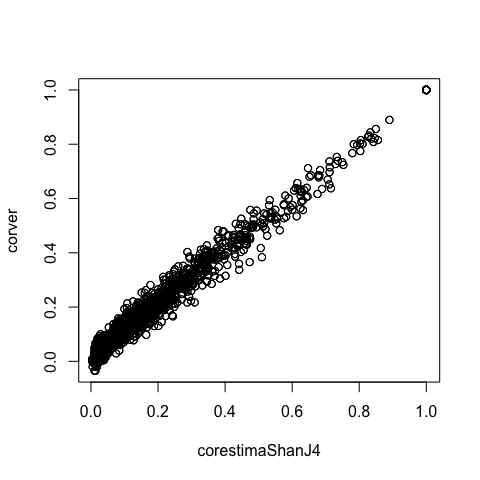}
        &
        \includegraphics[width=4.8cm]{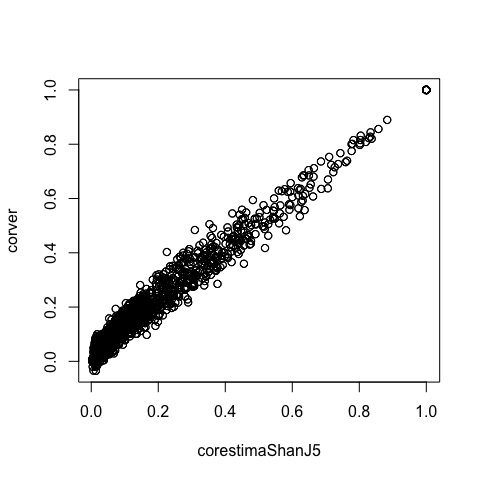}
    \end{tabular}
    \caption{Comparison of the estimated correlation matrix versus the true correlation matrix when $J = 3, 4, 5$ (from left to right) for quadratic deformation case using Mexican hat (upper) and Shannon (bottom) wavelets.}
\end{figure}

\begin{figure}[H]
\centering
    \begin{tabular}{lll}
        \includegraphics[width=4.8cm]{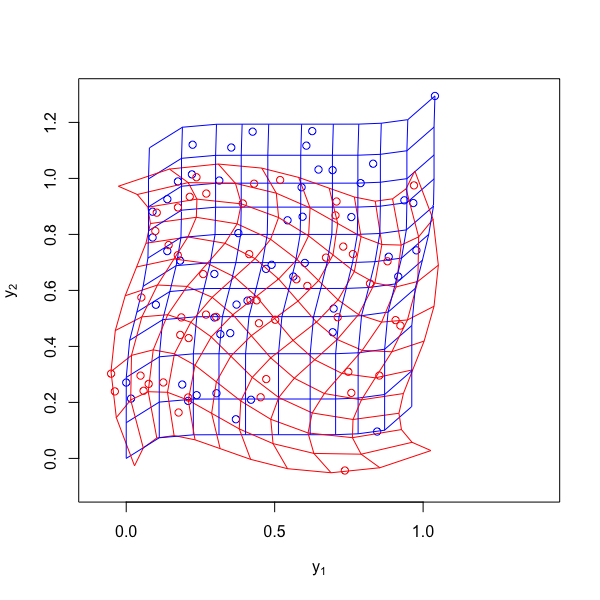}
        &
        \includegraphics[width=4.8cm]{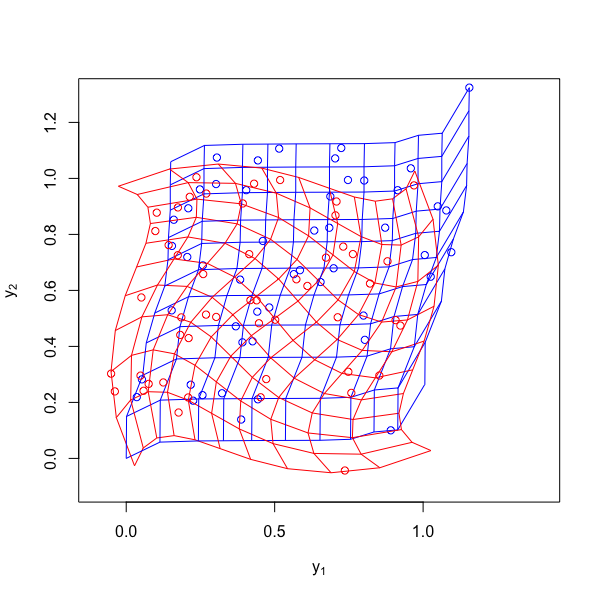}
        &
        \includegraphics[width=4.8cm]{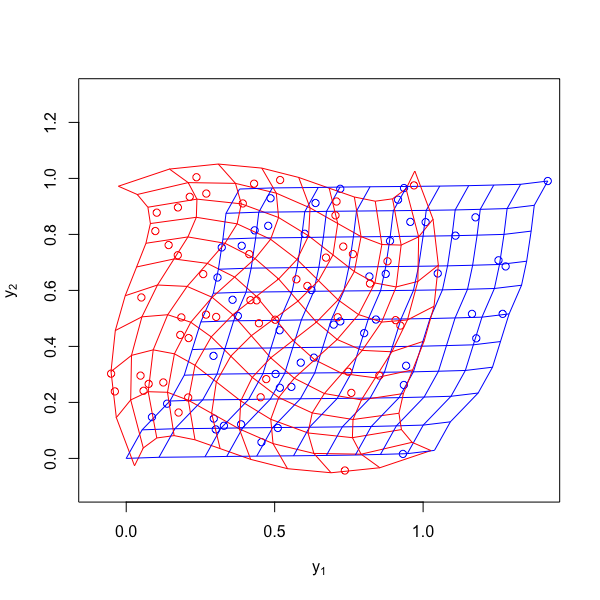}
        \\
        \includegraphics[width=4.8cm]{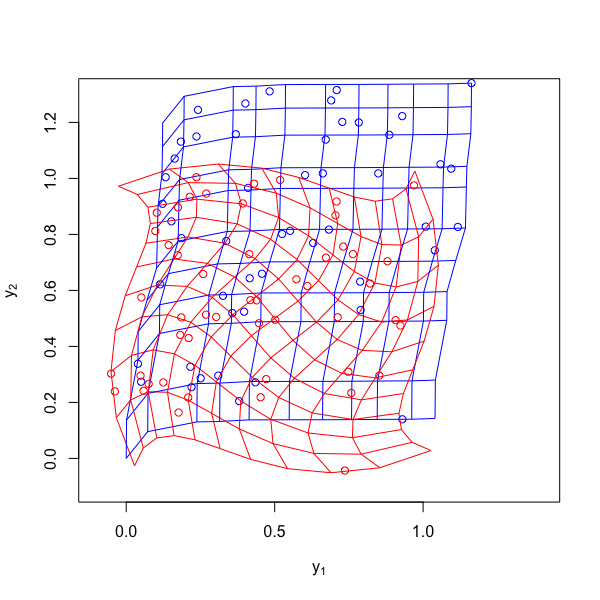}
        &
        \includegraphics[width=4.8cm]{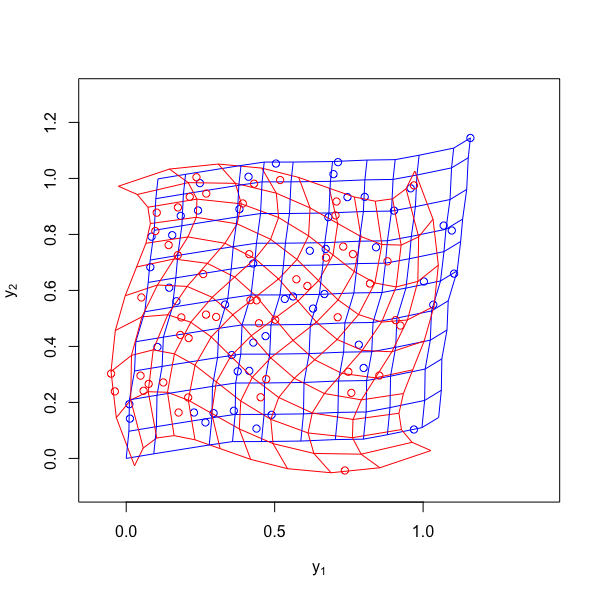}
        &
        \includegraphics[width=4.8cm]{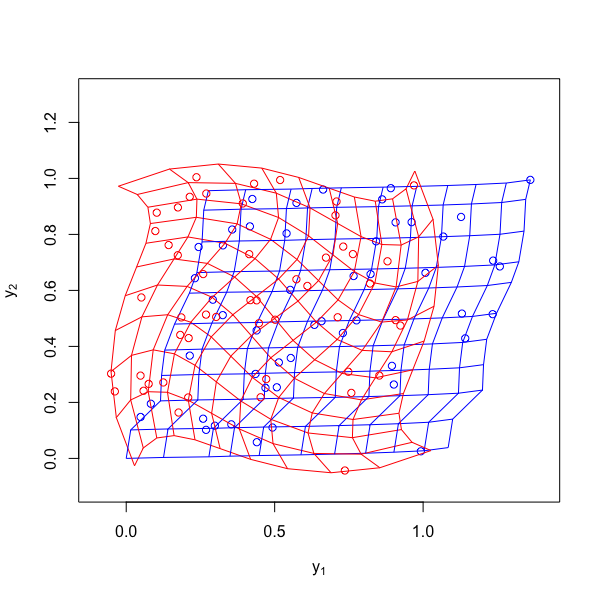}
    \end{tabular}
    \caption{Estimated deformation (in blue) when $J = 3, 4, 5$ (from left to right) for non-linear deformation case using Mexican hat (upper) and Shannon (bottom) wavelets.}
    \begin{tabular}{lll}
        \includegraphics[width=4.8cm]{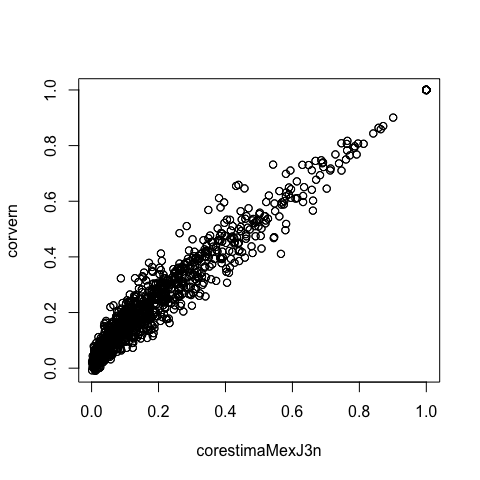}
        &
        \includegraphics[width=4.8cm]{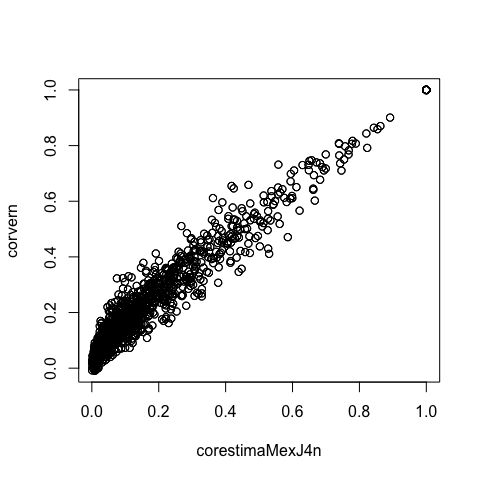}
        &
        \includegraphics[width=4.8cm]{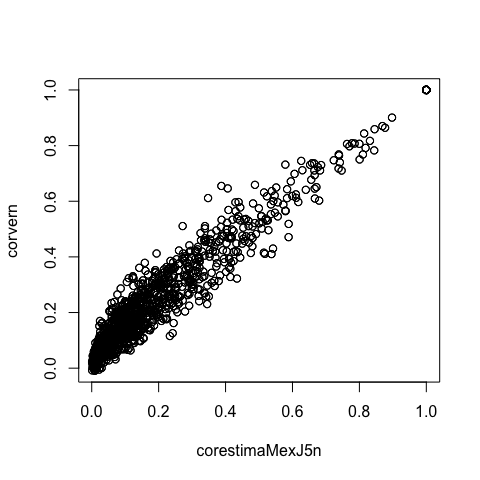}
        \\
        \includegraphics[width=4.8cm]{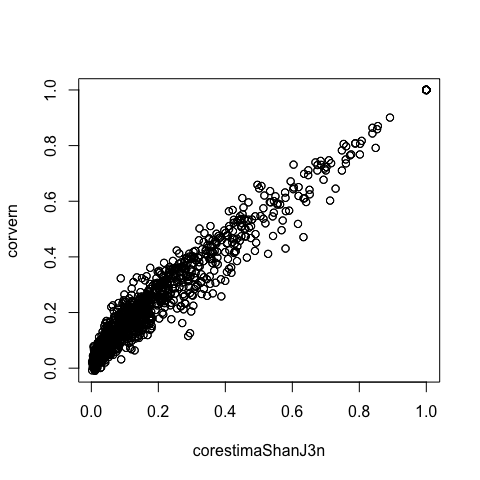}
        &
        \includegraphics[width=4.8cm]{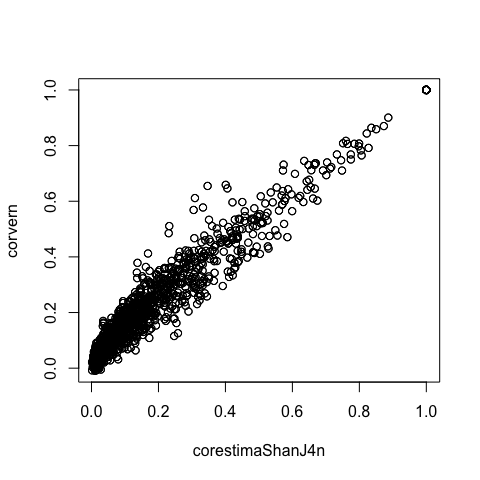}
        &
        \includegraphics[width=4.8cm]{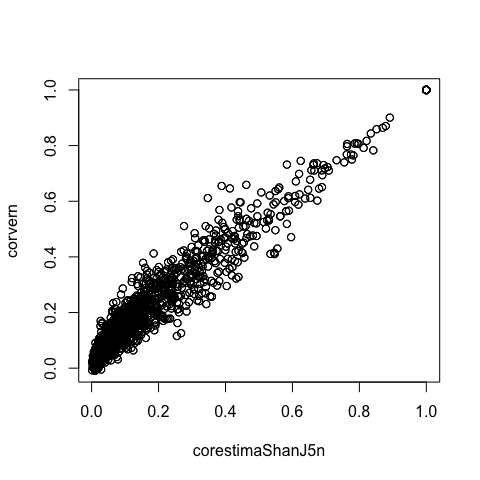}
    \end{tabular}
    \caption{Comparison of the estimated correlation matrix versus the true correlation matrix when $J = 3, 4, 5$ (from left to right) for non-linear deformation case using the Mexican hat (upper) and the Shannon (bottom) wavelets.}
\end{figure}

\begin{figure}[H]
\centering
    \begin{tabular}{lll}
        \includegraphics[width=4.8cm]{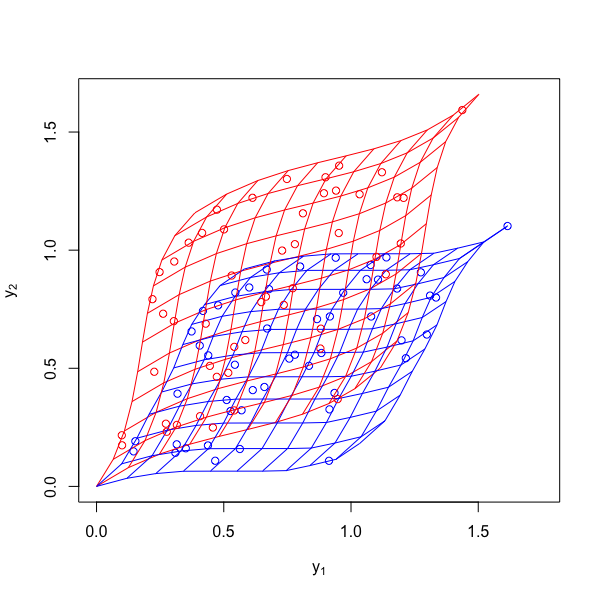}
        &
        \includegraphics[width=4.8cm]{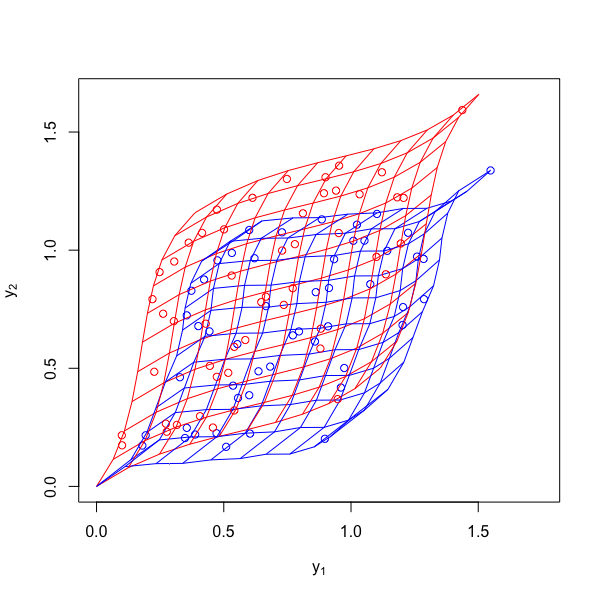}
        &
        \includegraphics[width=4.8cm]{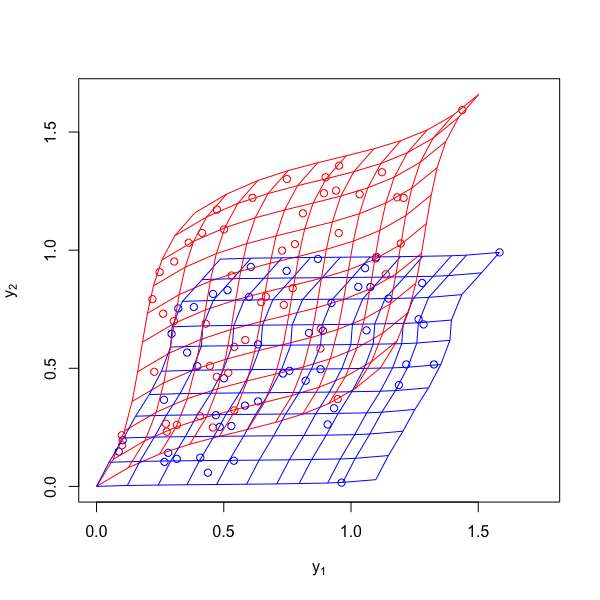}
        \\
        \includegraphics[width=4.8cm]{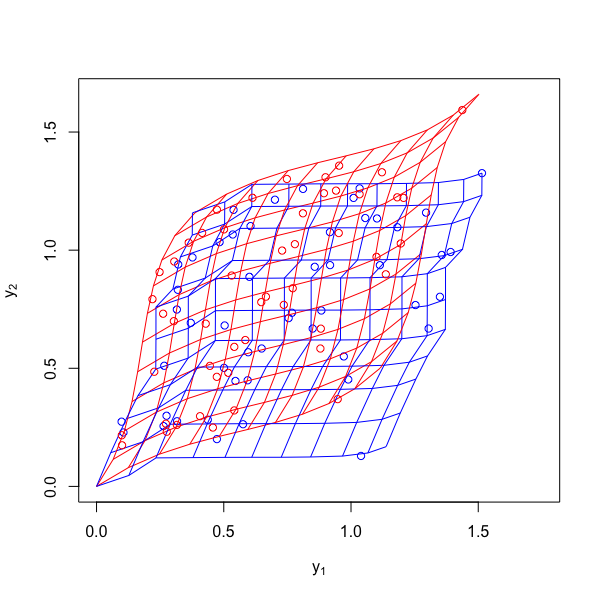}
        &
        \includegraphics[width=4.8cm]{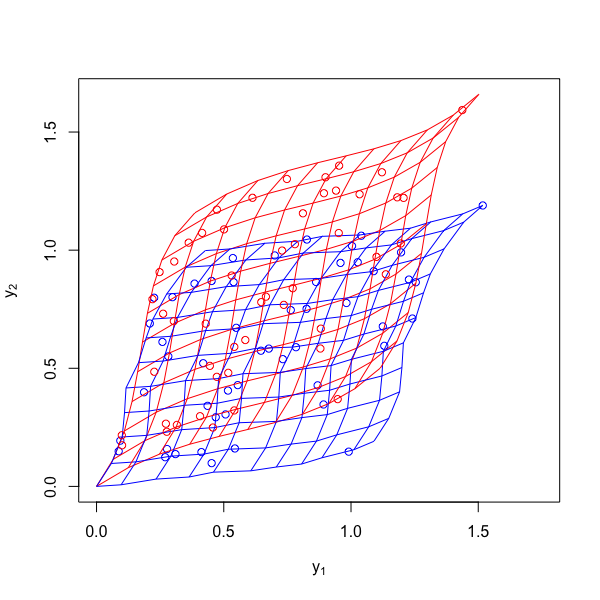}
        &
        \includegraphics[width=4.8cm]{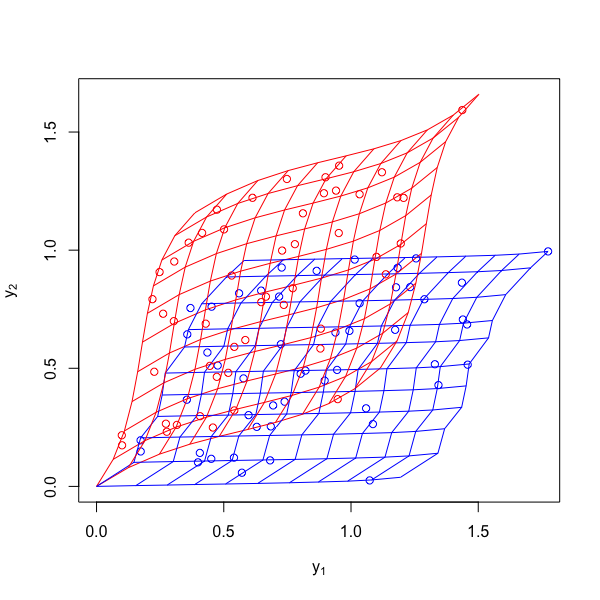}
    \end{tabular}
    \caption{Estimated deformation (in blue) when $J = 3, 4, 5$ (from left to right) for wavelet deformation case using the Mexican hat (upper) and the Shannon (bottom) wavelets.}
    \begin{tabular}{lll}
        \includegraphics[width=4.8cm]{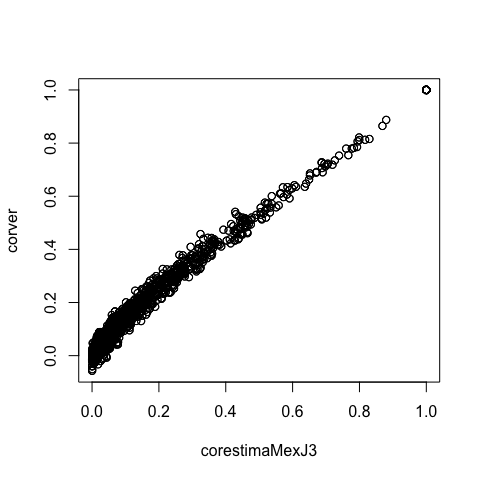}
        &
        \includegraphics[width=4.8cm]{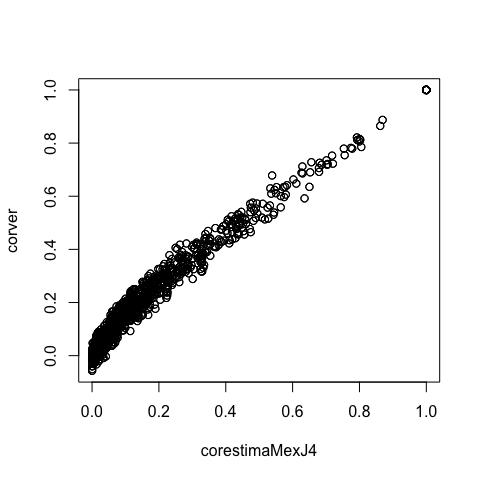}
        &
        \includegraphics[width=4.8cm]{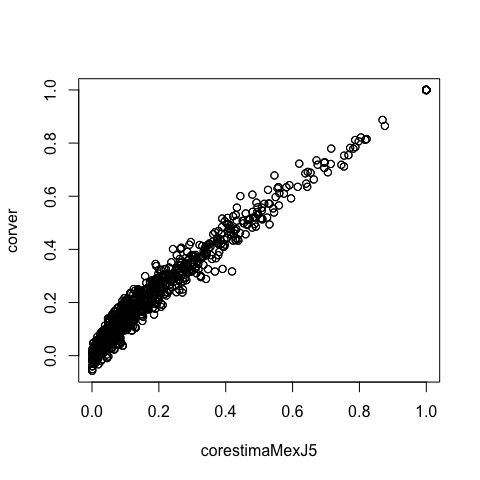}
        \\
        \includegraphics[width=4.8cm]{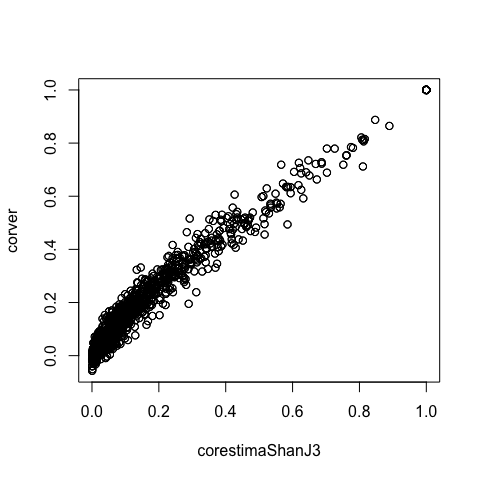}
        &
        \includegraphics[width=4.8cm]{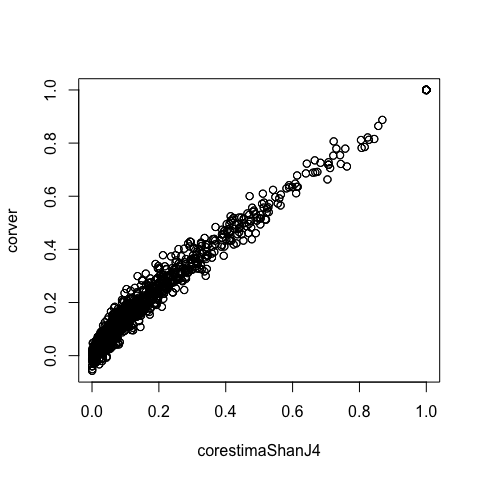}
        &
        \includegraphics[width=4.8cm]{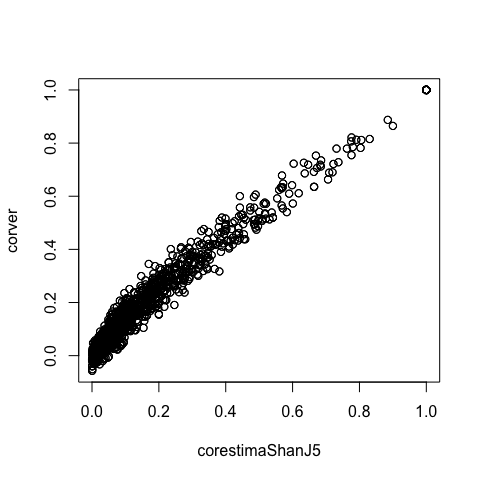}
    \end{tabular}
    \caption{Comparison of the estimated correlation matrix versus the true correlation matrix when $J = 3, 4, 5$ (from left to right) for wavelet deformation case using the Mexican hat (upper) and the Shannon (bottom) wavelets.}
\end{figure}

\section{Application}

The dataset used to illustrate the optimization procedure described in Section 3.4 provides historical daily maximum temperature records.
The data can be obtained directly from GHCN (Global Historical Climatology Network)-Daily, an integrated public database of NOAA (National Oceanic and Atmospheric Administration) using R package \textbf{rnoaa}.
The dataset deals with daily maximum temperature records (in tenths of degrees Celsius) from the Midwestern states of the USA. 
The region consists of 12 states: North Dakota, South Dakota, Illinois, Indiana, Iowa, Kansas, Michigan, Minnesota, Missouri, Nebraska, Ohio, and Wisconsin.
We use only climate monitoring stations containing no missing data between 1980 and 1999 (inclusive).
The 51 stations selected are shown in Figure 13. Figure 14 shows the maximum temperature recorded at the 4 sampling stations.

\begin{figure}[hp]
    \centering
    \includegraphics[width=9.5cm]{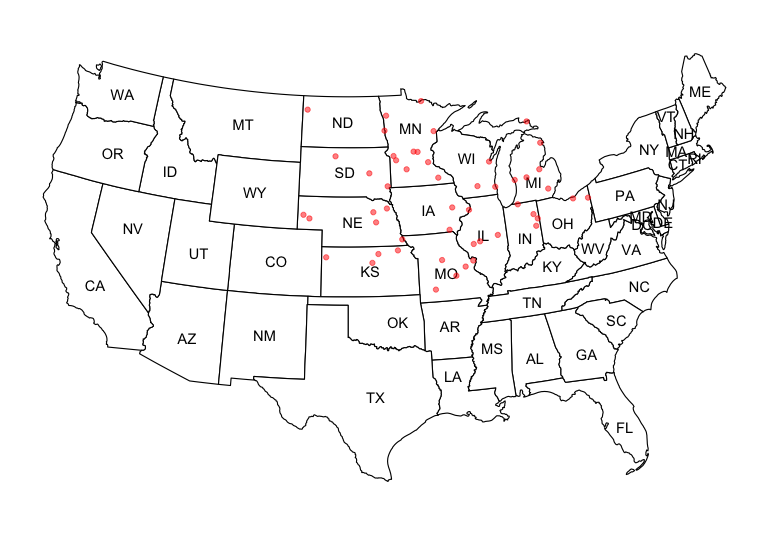} 
    \caption{Locations of the stations selected from Midwestern states of the USA.}
    \includegraphics[width=10cm]{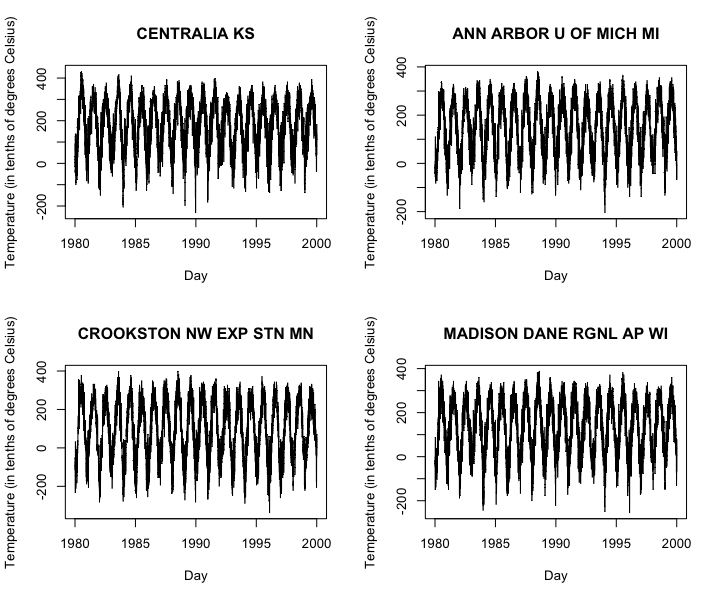}
    \caption{Maximum temperature recorded at the 4 sampling stations.}
\end{figure}

\newpage
The results were obtained from the optimization procedure described in Section 3.4 and deformation (22), (23) with J = 2, 3, 4 using Mexican hat and Shannon wavelets. And the covariance function used has the formula of equation (38). Figure 15 presents the estimated deformation and 
Table 2 shows the MSE's of the correlation matrix for the several fits. We can see that MSE are better for both Mexican hat and Shannon when J = 3. Same conclusions can be seen in the scatter plots (Figure 16). Note that the correlations are strong, maybe because the stations are from the same region.

\begin{table}[htb]
    \centering
    \caption{Estimated parameters and MSE of the correlation matrix for the several fits.}
    \begin{tabular}{ccccccc}
    \toprule
     &\multicolumn{3}{c}{Mexican hat} & \multicolumn{3}{c}{Shannon} \\ 
     \bottomrule
     &J = 2 & J = 3 & J = 4 & J = 2 & J = 3 & J = 4 \\\hline
    $\nu$   & 0,06438 & 0,06299 & 0,07400 & 0,06826 & 0,06871 & 0,06746 \\   
    $\theta$   & 5,11259 & 4,24264  & 5,86402 & 4,04539 & 4,11567 & 4,98565\\ 
    $\sigma^2_{\epsilon}$  & 0,00024 & 0,00016  & 0,00022 & 0,00015 & 0,00011 & 0,00019\\
    MSE & 0,00106 & 0,00103 & 0,00174 & 0,00146 & 0,00109 & 0,00122\\
    \bottomrule
\end{tabular}

\end{table}

\begin{figure}[H]
\centering
    \begin{tabular}{lll}
        \includegraphics[width=4.8cm]{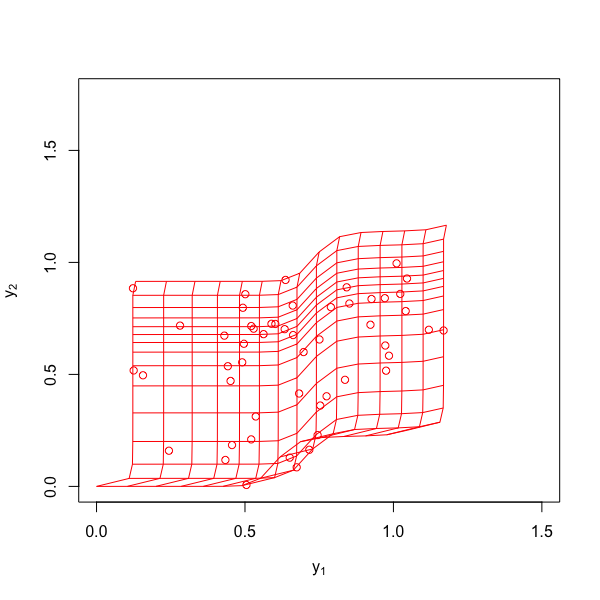}
        &
        \includegraphics[width=4.8cm]{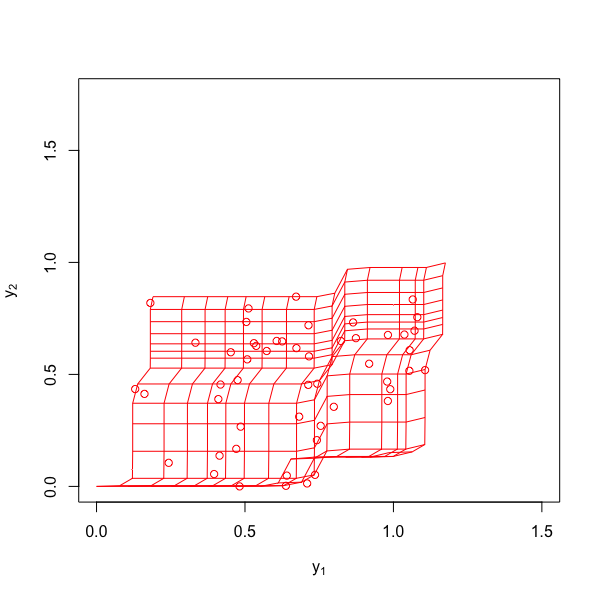}
        &
        \includegraphics[width=4.8cm]{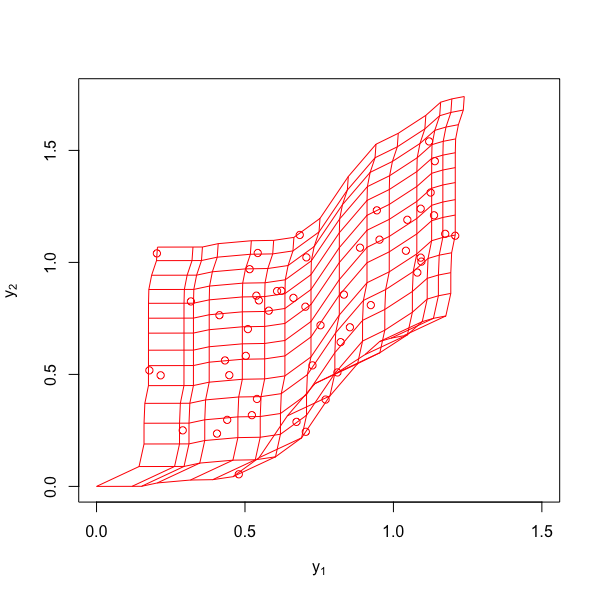}
        \\
        \includegraphics[width=4.8cm]{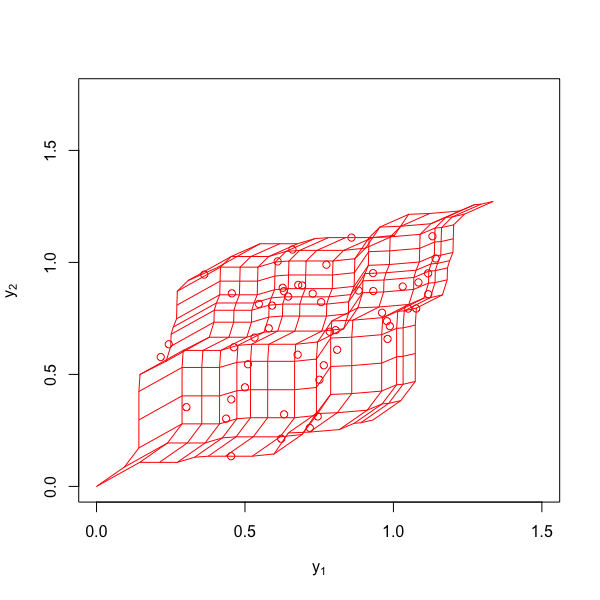}
        &
        \includegraphics[width=4.8cm]{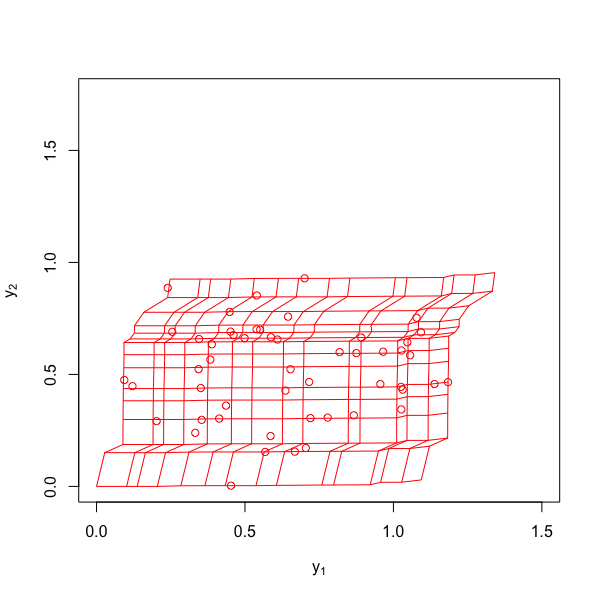}
        &
        \includegraphics[width=4.8cm]{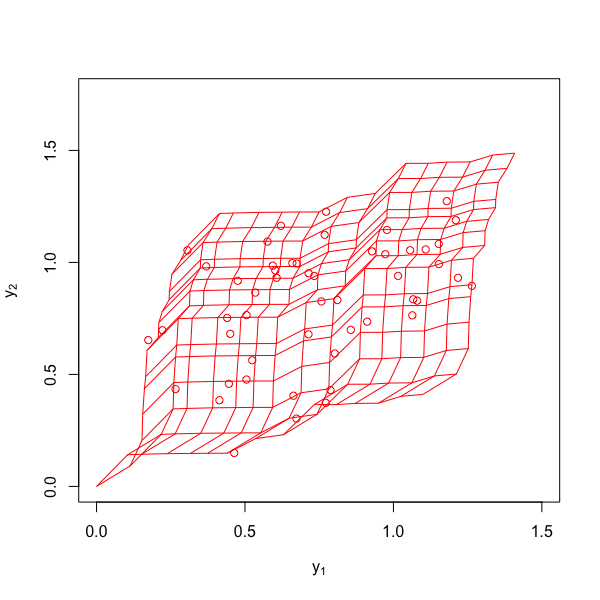}
    \end{tabular}
    \caption{Estimated deformation (in red) when $J = 2, 3, 4$ (from left to right) using Mexican Hat (upper) and Shannon (bottom) wavelets.}
\end{figure}

\begin{figure}[H]
\centering
    \begin{tabular}{lll}
        \includegraphics[width=4.8cm]{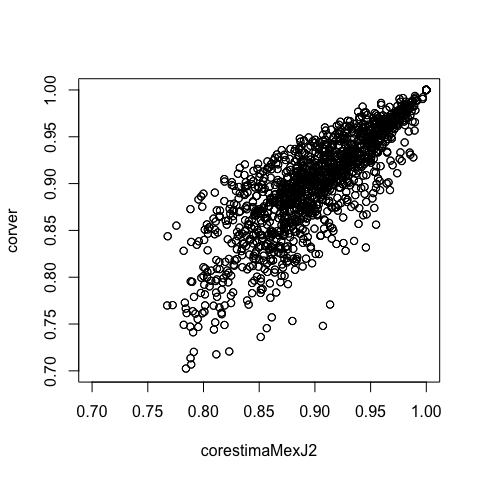}
        &
        \includegraphics[width=4.8cm]{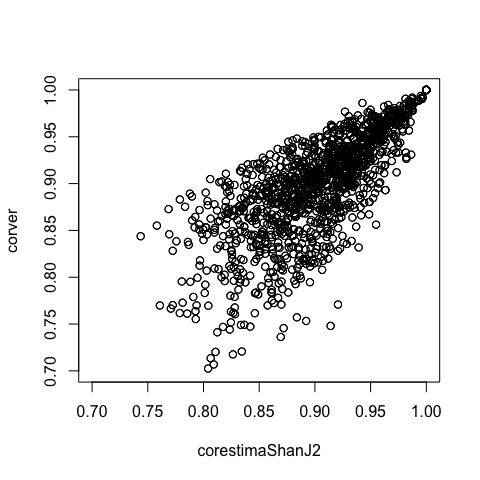}
        &
        \includegraphics[width=4.8cm]{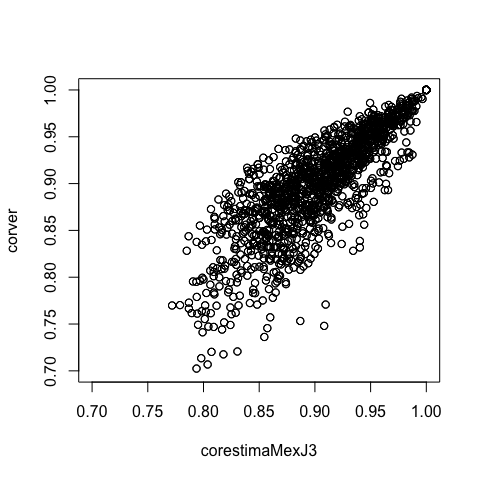}
        \\
        \includegraphics[width=4.8cm]{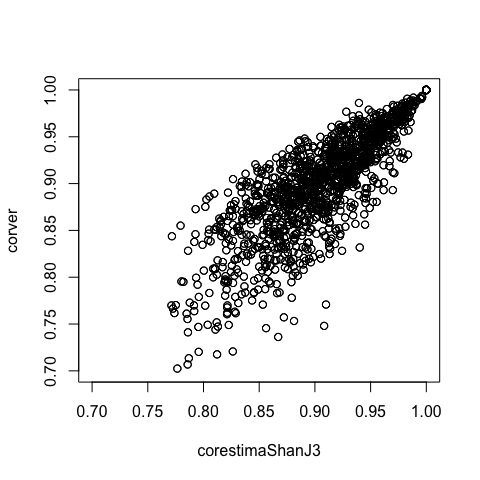}
        &
        \includegraphics[width=4.8cm]{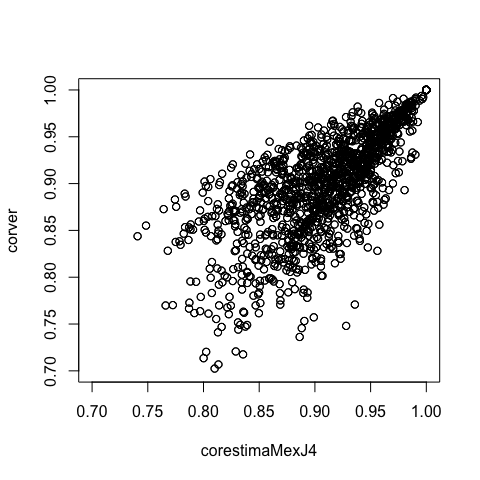}
        &
        \includegraphics[width=4.8cm]{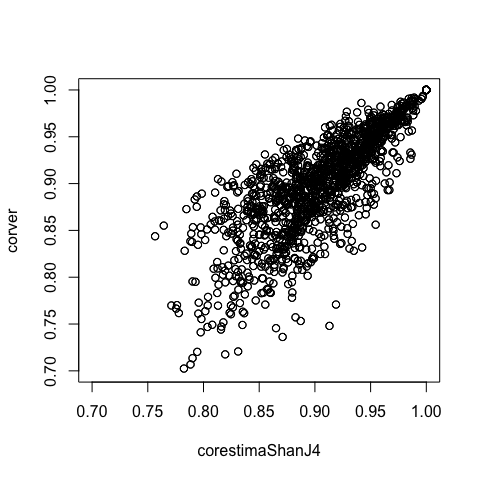}
    \end{tabular}
    \caption{Comparison of the estimated correlation matrix versus the true correlation matrix when $J = 2, 3, 4$ (from left to right) using Mexican Hat (upper) and Shannon (bottom) wavelets.}
\end{figure}

\section{Further comments}

In this paper, we have presented a deformation method based on the monotonic function and expansion of wavelet basis that guarantees the injetivity of the transformation. Some simulations and an application were performed with the proposed optimization. 

By the inverse function theorem, if the Jacobian determinant of $f$, denoted by $det(J(f))$, is non-zero, the $f$ is an injective function. The deformation proposed in this paper satisfies the condition that the Jacobian determinant is invertible. When $det(J(f)) > 0$, the $f$ is considered as a bijective function, that is, the orientation preserving of the function $f$ is guaranteed. Then, some methods that guarantee the bijectivity of the transformation under the stochastic model considered could be explored in future studies.

\section*{Acknowledgements}

Yangyang Chen acknowledges the support of FAPESP, through grant 2019/05917-6. Pedro Alberto Morettin, Ronaldo Dias and Chang Chiann acknowledge the partial support of FAPESP grant 2018/04654-9.

\section*{References}
\addcontentsline{toc}{chapter}{\protect\numberline{}References}%

Choi, Y. and Lee, S. (2000) Injectivity conditions of 2D and 3D uniform cubic B-spline functions, \textit{Graphical Models}, vol.62, Issue 6, 411–427.

Chun, S. Y. and Fessler, J. A. (2009) A simple regularizer for B-spline nonrigid image restration that encourages local invertibility, \textit{IEEE journal of selected topics in signal processing} \textbf{3}(1): 159-169.

Cressie, N. and Wikle, C. K. (2011) \textit{Statistic for Spatio-Temporal Data}, New York: Wiley.

Damian, D., Sampson, P. and Guttorp, P. (2001) Bayesian estimation of semiparametric nonstationary spatial covariance structures, \textit{Environmentrics} \textbf{12}(2):161-178.

Guttorp, P., Meiring, W. and Sampson, P. D. (1994) A space-time analysis of ground-level ozone data, \textit{Environmetrics} \textbf{5}(3):241-254.

Hardle, W., Kerkyacharian, G., Picard, D., Tsybakov, A. (1997) \textit{Wavelets Approximations and Statistical Applications}, New York: Springer.

Hastie, T.J. and Tibshirani, R.J. (1990) \textit{Generalized Additive Models}, Chapman and Hall, New York.

Kim, J. (2004) Intensity Based Image Registration Using Robust Similarity Measure and Constrained Optimization: Applications for Radiation Therapy, Ph.D. dissertation, University of Michigan, Ann Arbor, MI.

Musse, O. Heitz, F. and Armspach, J. (2001) Topology preserving deformable image matching using constrained hierarchical parametric models, \textit{IEEE Trans. Image Process.}, vol. 10, Inssue 7, 1081–1093.

Ramsay, J. O. (1998) Estimating smooth monotone functions. \textit{Journal of Royal Statistical Society}, Series B, \textbf{60}(2), 365-375.

Ramsay, J. O. and Silverman, B. W. (2006) \textit{Functional Data Analysis}, 2nd Edition, New York: Springer.

Sampson, P. and Guttorp, P. (1992) Nonparametric estimation of nonstationary spatial covariance structure, \textit{Jounal of the American Statistical Association} \textbf{87}(417): 108-119.

Schmidt, A. M. and O'Hagan, A. (2003) Bayesian inference for non-stationary spatial covariance structure via spatial deformations. \textit{Journal of the Royal Statistical Society}, Series B, \textbf{65}(3), 743-758.

\end{document}